  \documentstyle[12pt]{article}

\def\journal#1, #2, #3, #4 { {\sl #1~}{\bf #2~} (#3)  #4 }

\def\prd{\journal Phys. Rev. D, }

\def\prl{\journal Phys. Rev. Lett., }

\def\cmp{\journal Comm. Math. Phys., }

\def\np{\journal Nucl. Phys., }

\def\pl{\journal Phys. Lett., }

\def\annp{\journal Ann. Phys. (N.Y.), }

\def\ijmp{\journal Int. J. Mod. Phys., }

\catcode`\@=11
\def\marginnote#1{}
\newcount\hour
\newcount\minute
\newtoks\amorpm
\hour=\time\divide\hour by60
\minute=\time{\multiply\hour by60 \global\advance\minute
by-\hour}\edef\standardtime{{\ifnum\hour<12
\global\amorpm={am}%
        \else\global\amorpm={pm}\advance\hour by-12 \fi
        \ifnum\hour=0 \hour=12 \fi
        \number\hour:\ifnum\minute<10
0\fi\number\minute\the\amorpm}}
\edef\militarytime{\number\hour:\ifnum\minute<10
0\fi\number\minute}

\def\draftlabel#1{{\@bsphack\if@filesw {\let\thepage\relax
   \xdef\@gtempa{\write\@auxout{\string
      \newlabel{#1}{{\@currentlabel}{\thepage}}}}
Ebenen
}\@gtempa
   \if@nobreak \ifvmode\nobreak\fi\fi\fi\@esphack}
        \gdef\@eqnlabel{#1}}
\def\@eqnlabel{}
\def\
Ebenen
@vacuum{}
\def\draftmarginnote#1{\marginpar{\raggedright\scriptsize\tt#1}}
\def\draft{\oddsidemargin -.5truein
        \def\@oddfoot{\sl preliminary draft \hfil
        \rm\thepage\hfil\sl\today\quad\militarytime}
        \let\@evenfoot\@oddfoot \overfullrule 3pt
        \let\label=\draftlabel
        \let\marginnote=\draftmarginnote

\def\@eqnnum{(\theequation)\rlap{\kern\marginparsep\tt\@eqnlabel}%
\global\let\@eqnlabel\@vacuum}  }

\def\numberbysection{\@addtoreset{equation}{section}
        \def\theequation{\thesection.\arabic{equation}}}

\def\underline#1{\relax\ifmmode\@@underline#1\else
 $\@@underline{\hbox{#1}}$\relax\fi}

\catcode`@=12
\relax
\numberbysection

\def\beq{\begin{equation}}
\def\eeq{\end{equation}}
\def\beqa{\begin{eqnarray}}
\def\eeqa{\end{eqnarray}}
 \def\nnn{\nonumber \\}

\def\hhat{{\widehat h}}
\def\lfloorhat{{\hat \lfloor}}
\def\rfloorhat{{\hat \rfloor}}

\def\Jhat{{\widehat J}}

\def\Je{J^e{}}

\def\Jehat{{\Jhat^e}{}}

\def\kappahat{{\widehat \kappa}}

\def\mhat{{\widehat m}}
\def\nhat{{\widehat n}}

\def\Vhat{{\widehat V}}
\def\Uhat{{\widehat U}}

\def\xhat{{\widehat x}}

\def\xib{{\overline \xi}}

\def\Vhat{{\widehat V}}

\def\muhat{{\widehat \mu}}

\def\qhat{{\widehat q}}

\def\varpihat{{\widehat \varpi}}

\def\varpib{{\overline \varpi}}

\def\kappab{{\overline \kappa}}

\def\mhat{{\widehat m}}
\def\nhat{{\widehat n}}

\def\Shat{{\widehat S}}

\def\xb{{\bar x}}
\def\Jb{{\bar J}}
\def\Mb{{\bar M}}
\def\sb{{\bar s}}
\def\Sb{{\overline  S}}
\def\zb{{\bar z}}

\def\mb{{\bar m}}

\def\Shatb{{\widehat {\overline S}}}

\def\etab{\bar \eta}
\def\nub{\bar \nu}
\def\Mc{{\cal M}}
\def\Mcb{{\bar {\cal M}}}

\def\Jgen#1 {  {\underline J_{#1}} }
\def\Kgen#1 {  {\underline K_{#1}} }
\def\Jgenp#1 #2 {(J_{#1}+{#2},\Jhat_{#1})}
\def\Jgenm#1 #2 {(J_{#1}-{#2},\Jhat_{#1})}
\def\Jg#1 {J_{#1},\Jhat_{#1}}
\def\Jgp#1 #2 {J_{#1}+{#2},\Jhat_{#1}}
\def\Mgen#1 {{\underline M_{#1}}}
\def\mgen#1 {{\underline m_{#1}}}
\def\xgen{{\underline x}}
\def\ms{m^\circ{}}
\def\msone{m_1^\circ}
\def\mstwo{m_2^\circ}
\def\mshat{ {\mhat^\circ}{}}
\def\msonehat{{\mhat_1^\circ}{}}
\def\mstwohat{{\mhat_2^\circ}{}}
\def\mssum{m_{12}^\circ{}}
\def\mssumhat{{\mhat_{12}^\circ}{}}
\def\Vt{{\widetilde V}}

\def\Vtb{\overline{\Vt}}

\def\fin{\end{document}}

\def\Jge{{\underline J}}
\def\mge{{\underline m}}

\def\Jgen#1 {  {\underline J_{#1}} }
\def\Jgenp#1 #2 {(J_{#1}+{#2},\Jhat_{#1})}
\def\Jgenm#1 #2 {(J_{#1}-{#2},\Jhat_{#1})}
\def\Jg#1 {J_{#1},\Jhat_{#1}}
\def\Jgp#1 #2 {J_{#1}+{#2},\Jhat_{#1}}
\def\Mgen#1 {{\underline M_{#1}}}

\def\fusV#1,#2,#3,#4,#5,#6 {f_V(
\Jgen{#1} ,
\Jgen{#2} ,
\Jgen{#3} ,
\Jgen{#4} ,
\Jgen{#5} ,
\Jgen{#6} )}

\def\sixjxi#1,#2,#3,#4,#5,#6
{{\left\{\left . \!\! \,^{#1}_{#2}\,^{#3}_{#4}
\right | \!\, ^{#5}_{#6}\right\}}}

\def\gaghat{{\hat {\bigl \{}}}
\def\gadhat{{\hat {\bigr \}}}}
\def\Gaghat{{\hat {\Bigl \{}}}
\def\Gadhat{{\hat {\Bigr \}}}}

\def\bverthat{{\hat {\Bigl \vert}}}

\begin{document}
\tolerance 2000
\hbadness 2000
\begin{titlepage}

\nopagebreak

\vglue 3  true cm
\begin{center}
{\large \bf
QUANTUM GROUP STRUCTURE AND LOCAL\\
\medskip
FIELDS IN THE ALGEBRAIC APPROACH\\
\medskip
TO 2D GRAVITY}\\
\vglue 1.5 true cm
\medskip
{\bf Jens SCHNITTGER}\\
\medskip
{\footnotesize CERN-TH, 1211 Geneva 23, Switzerland}.\\
\end{center}
\vfill
\vglue 1 true cm
\begin{abstract}
\baselineskip .4 true cm
{\footnotesize
\noindent
This review contains a summary of work by J.-L. Gervais and the
author
on the operator approach to 2d gravity. Special emphasis is placed
on the construction of local observables -the Liouville exponentials
and
the Liouville field itself - and the underlying algebra of chiral
vertex
operators. The double quantum group structure arising from the
presence of two
screening charges is discussed and the generalized algebra and field
operators
are derived.
In the last part, we show that our construction gives rise to a
natural definition of a quantum tau function, which is a
noncommutative
version of the classical group-theoretic representation of the
Liouville
fields  by Leznov and Saveliev.
}
\end{abstract}
\vfill
\end{titlepage}
\section{Introduction}
Two-dimensional gravity has been the subject of intensive study in
recent
years. The  approaches which were employed cover a wide spectrum,
ranging from topological theory to
matrix models, and this lead to the discovery of fruitful relations
and cross-connections. Each of the methods has its own particular
merits and
shortcomings, and so the preference of one over the other depends to
some
extent on the nature of the question we want to address. The
traditional
virtue of the operator approach is to provide insight into the
{\underline{local}}
structure of the theory, i.e. into operator algebra and  correlation
functions, instead of fusion rules and amplitudes only.
This may not be so important if we are only interested in the
calculation of
noncritical string scattering amplitudes, but is evidently crucial if
we want
to understand the structure of 2d gravity/Liouville theory as a
conformal
field theory. The most spectacular success produced so far by the
algebraic
method was the discovery of the underlying quantum group
symmetry\cite{B}
\cite{G1}\cite{G3}, which had remained completely invisible in the
other
approaches. Until recent times, the main technical tool in this
framework
 was the detailed study of the Virasoro null vector equations and
their monodromy properties\cite{B}-\cite{CGR2}, which in turn are
related to
representations of the $U_q(sl(2))$ quantum group with positive
half-integer
spins\cite{CGR1}\cite{CGR2}.
More precisely, one finds that a subset of the observables of the
theory - the
positive integer powers of the inverse square root of the metric -
are described by chiral vertex operators which are degenerate fields
in the sense of BPZ\cite{BPZ}.
However, the study of representations with positive half-integer
spins
 does not
by far  answer all the physical questions we want to ask
about 2D gravity.
For instance, the coupling of minimal matter to gravity
requires negative half-integer spins, and modular invariance in
the strong coupling version of the theory
forces us to consider operators with
quantum-group spins which are rational,
but  not halves of integers\cite{GR}. Of course the full operator
content of the theory, and in particular the possibility of
defining the Liouville field itself,
is explored only if we can define the Liouville exponential with
continuous $J$, so that we may let $\Phi =-{d\over dJ}|_{J
=0} \exp (-J\alpha_- \Phi)/ \alpha_- $.
The basic difficulty in
going away from positive half-integer spins is that one no longer
deals
with
degenerate fields satisfying null-vector equations. We will show how
to
solve this problem
by introducing an operator Coulomb-gas realization of the chiral
vertex operators. This allows  us to
establish
their exchange algebra , which is the essential ingredient
for the reconstruction of local
observables.
The crucial observation in this context is
that the braiding problem, due to its essentially  topological
nature,
can be
represented in terms of simple quantum-mechanical variables.

Another interesting aspect of Liouville theory is its connection with
the general theory of integrable systems. As the simplest in the
sequence of
$A_n$ Toda theories,
it should possess both a description and a multi-time generalization
in terms of  Toda lattice $\tau$ functions and their reductions.
Though this
is completely understood classically, even the very concept of a
$\tau$ function is still under discussion on the quantum
level\cite{KS}\cite{SMI}\cite{GKLMM}
\cite{KM}\cite{MMV}. As a contribution to this discussion,
 we present an argument
why our quantum  exponentials should possess an interpretation
as quantum $\tau$ functions. The analysis is based on the (classical)
group-theoretical representation of the Liouville exponentials by
Leznov and Saveliev\cite{LS}, which is known to have the properties
of a $\tau$ function.
Due to the explicit control furnished by the operator approach over
the
quantization of all the defining elements of this representation, we
are able
to write the natural quantum equivalent of the Leznov-Saveliev $\tau$
function.

This review is organized as follows. In section 2 we recall some
elementary
background material. Section 3 is devoted to the case of a single
screening charge, where the quantum group structure is known to be
the
standard $U_q(sl(2))$. We work with Coulomb-gas type operators
which have a diagonal monodromy, i.e. well-defined periodicity
properties,
instead of  explicitly quantum group covariant operators.
This picture has the advantage that the chiral exchange algebra can
be derived using only free field methods, as we will show. The fusion
algebra (operator product) is not as easily accessible by this
method,
but can be obtained from the braiding algebra if one assumes the
validity
of the general Moore-Seiberg\cite{MS} proportionality relation
between fusing
and braiding matrices.
The control over the chiral algebra then allows us  to construct
Liouville
exponentials for arbitrary $J$, and we give an expression for the
Liouville
field itself. The validity of the canonical equal-time commutation
relations  as well as the quantum equations of motion is
demonstrated.
We discuss the periodicity properties of our definition
 of the Liouville exponentials
resp. of the Liouville field and their connection with the presence
of singularities in the elliptic sector. The preservation on the
quantum level
of the symmetry under the exchange of the two equivalent B\"acklund
free
fields\cite{GN3} -which has the meaning of a Weyl reflection- is
derived for half-integer spins.
In section 4, we consider both screening charges
together. The corresponding quantum group structure
$U_q(sl(2))\odot U_{\qhat}(sl(2))$ is known\cite{CGR1} to be given by
a
sort of graded tensor product  in the degenerate
case, where the primary fields of   spins $J$ and $\Jhat$ are of the
type
$(2\Jhat+1, 2J+1)$ in the BPZ classification. This is because
the fields of type
$(1, 2J+1)$ field commute with those of type  $(2\Jhat+1, 1)$ up
to a simple phase.
For continous spins, however, the situation is  different,
and the notion of two independent spins $J$ and $\Jhat$ becomes
meaningless. The latter two unite into a single parameter
$J^e :=J+\Jhat \pi/h$ ($h$ is defined by $q=\exp (ih)$ as
usual), and a new factorization property of the fusion
and braiding matrices in terms of the two screening charges emerges.
The new $U_{q\qhat}(sl(2))$ structure can be described in terms
of ``vectorial'' 6j symbols which have been shown in \cite{GR} to
satisfy all of the polynomial equations of Moore and
Seiberg\cite{MS}.
The corresponding representations have one spin but two magnetic
quantum
numbers, corresponding to the two types of screening charges.

In the last section, we consider the integrability structure of the
theory
from the $\tau$ function point of view.
We start from the group-theoretical representation of the classical
Liouville exponentials by Leznov and Saveliev\cite{LS}, which can be
understood as a classical tau function and in particular fulfills
bilinear equations of the Hirota type. The quantum group covariant
quantization scheme described here provides a natural prescription
for the
 proper quantization of the Leznov-Saveliev formula, such that its
algebraic
and group theoretical properties are just the obvious q-deformations
of those
of the classical case. For this purpose it is  most natural to work
with
an explicitly quantum group covariant operator
basis\cite{G1}\cite{G3}\cite{CGR2}, where fusion
and braiding are given in terms of $3j$ symbols and  universal
$R$-matrix.

\section{Some background material}
\label{self-contained}
Let us  rederive  some background material about
Liouville theory, in order to introduce  the coming discussion.
The solutions of the classical Liouville dynamics, which is
described by the action
\begin{equation}
S={1\over 8\pi}\int d\tau d\sigma  \{ (\partial_\tau \Phi )^2
-(\partial_\sigma \Phi )^2 -\mu^2 e^{2 \sqrt{\gamma} \Phi} \},
\label{2.1}
\end{equation}
are given by \hfill
\begin{equation}
2\sqrt{\gamma} \Phi =
\ln \left [{8\over \mu^2\sqrt{\gamma}}
{A'(u)B'(v)\over  (A(u)-B(v))^2 }\right ]
\qquad , \qquad u=\tau +\sigma, \quad v=\tau -\sigma
\label{2.2}
\end{equation}
with $A$ and $B$ arbitrary functions, and $\sigma \in [0,2\pi ]$.
The coupling constant is noted $\gamma$.
We have redefined $\Phi \to 2\sqrt{\gamma}\Phi$ in order to agree
with
the classical limit of
standard quantum normalizations, where $2\sqrt{\gamma}$ is the limit
of
the
screening charge $\alpha_-$.
Eq.\ref{2.2} is invariant under the projective transformations
\begin{equation}
A\to {aA+b\over cA+d} \qquad B\to {aB+b\over cB+d} \qquad ,
\label{2.3}
\end{equation}
$a,b,c,d$ complex, which on the quantum level gives rise to the
$U_q(sl(2))$ quantum group symmetry. Introducing the chiral
fields\footnote{In this formula, contrary to the rest of the
article, we of course  use ordinary binomial coefficients.}
\begin{eqnarray}
f_m^{(J)}&=& \sqrt{\scriptstyle {2J \choose J+m}}
(A'^{-1/2})^{J-m} (AA'^{-1/2})^{J+m},\nnn
\bar f_m^{(J)}&=&\sqrt{\scriptstyle {2J \choose J+m}}
(B'^{-1/2})^{J+m} (BB'^{-1/2})^{J-m},
\label{2.4}
\end{eqnarray}
we can write the Liouville exponentials as
\begin{equation}
e^{-2J \sqrt{\gamma}\Phi} =\left ({\mu^2 \sqrt{\gamma}\over 8}\right
)^J
\sum_{m=-J}^{J}  (-1)^{J+m}
f_m^{(J)}\bar f_m^{(J)}
\label{2.5}
\end{equation}
for any positive half-integer $J$. It is easy to verify that,
when $A$ and $B$ undergo  the M\"obius transformation
Eq.\ref{2.3},   the functions
$f_m^{(J)},\bar f_m^{(J)}$ transform as standard
 (finite-dimensional)
spin $J$ representations of $sl(2,\bf{C})$, and Eq.\ref{2.5} is the
corresponding singlet.
For general  $J$, Eq.\ref{2.5}
is still valid formally, with the $m-$sum extending to $+\infty$
or  to $-\infty$, depending on whether we work with the
semi-infinite  representations with $J+m$ positive integer,
or $J-m$ positive integer. In practice, Eq.\ref{2.5} is
an expansion  in $(A/B)^m$, and
highest-weight representations, with  $-\infty\leq m\leq J$
(resp. lowest-weight representations, with $-J \leq m\leq \infty$)
will
give a convergent expansion for $|A|> |B|$ (resp. $|B| > |A|$).
Both choices should
 represent the same function, since
they are just related by the particular $sl(2)$-transformation
\begin{equation}
A\to -1/A  \qquad B\to -1/B \quad ,
\label{2.6}
\end{equation}
which sends $m \to -m$ in $f_m^{(J)}, \bar
f_m^{(J)}$.\footnote{Actually
the situation is somewhat more subtle for singular solutions,
as we will see later.}  Of course,
in the case of positive half-integer $J$, this amounts only to a
trivial permutation
of terms in the sum Eq.\ref{2.5}. For continous $J$, however, the
highest
resp. lowest weight representations are representations only of the
algebra
but not of the group, due to the multivaluedness of the $f_m^{(J)}$
under the
group operations Eq.\ref{2.3}. Consequently, the
transformation Eq.\ref{2.6} exchanges highest and lowest
weight representations. From the general point of view of Toda
theory,
it can be regarded as representing the Weyl group
symmetry\cite{ANPS}.

Periodicity of $\Phi$ implies that $A$ and $B$ must be periodic up to
a projective transformation, which is called the monodromy matrix.
In the elliptic and hyperbolic sectors of the theory, we can always
pick a representative of the equivalence class defined by
Eq.\ref{2.3}
such that the monodromy matrix is diagonal, i.e. such that $A$ and
$B$ are periodic up to a multiplicative constant. In fact, there are
precisely two such representatives, related by Eq.\ref{2.6} which can
thus be viewed in this context as a kind of residual symmetry.
It is then possible to define two equivalent sets of chiral free
fields by\footnote{Compared to ref.\cite{GS2}, we have changed the
notation
by replacing $\phi_1 \to \vartheta_1,  \  \phi_2 \to \vartheta_2,
\ -\bar\phi_1 \to \bar\vartheta_2,
\ -\bar\phi_2 \to \bar\vartheta_1$
.}
\begin{eqnarray}
\sqrt{{\gamma }}\vartheta_1(u):=
\ln A'^{-1/2}(u),&  \quad  \sqrt{{\gamma }}
\bar\vartheta_1(v):=
\ln BB'^{-1/2}(v), \nnn
\sqrt{{\gamma}}\vartheta_2(u):=\ln AA'^{-1/2}(u), &
\quad \sqrt{{\gamma}} \bar \vartheta_2(v):=
\ln B'^{-1/2}(v)
\label{freefields}
\end{eqnarray}
Indeed, one may show that the canonical Poisson brackets
 of Liouville theory
give the following free-field Poisson bracket relations
\begin{equation}
\Bigl \{\vartheta'_1(\sigma_1),\vartheta'_1(\sigma_2)
\Bigr \}_{\hbox {\footnotesize P.B.}}=
\Bigl \{\vartheta'_2(\sigma_1),\vartheta'_2(\sigma_2)
\Bigr \}_{\hbox {\footnotesize P.B.}}
=2\pi \,  \delta'(\sigma_1-\sigma_2)
\label{2.8}
\end{equation}
with similar relations for the bar components.
The explicit mode expansions in terms of zero modes and oscillators
are given by
\begin{eqnarray}
\phantom{1234567890}&\vartheta_j(u)=q^{(j)}_0+ p^{(j)}_0 u+
i\sum_{n\not= 0}e^{-in u}\, p_n^{(j)}\bigl / n,
\quad \phantom{j=1,\> 2}  \phantom{1234567890}\cr
\phantom{1234567890}&\bar\vartheta_j(v)=\bar q^{(j)}_0+\bar
p_0^{(j)}v+
i\sum_{n\not= 0}e^{-in v}\, \bar p_n^{(j)}\bigl / n,\quad j=1,\> 2
\phantom{1234567890}\cr
\label{1.8}
\end{eqnarray}
{}From Eq.\ref{1.8} we see that
the periodicity properties of the $A$ and $B$ fields can
be parametrized by the zero mode momenta e.g. of $\vartheta_1$:
\beq
A(u+2\pi)=e^{-4\pi p_0^{(1)}\sqrt{\gamma}} A(u), \quad
B(v+2\pi)=e^{+4\pi \bar p_0^{(1)}\sqrt{\gamma}} B(v)
\label{2.9}
\eeq
The complete symmetry of the treatment of the theory
 under the exchange of
$\vartheta_1,\bar \vartheta_1$
and
$\vartheta_2,\bar \vartheta_2$ even on the quantum level  is  the
hallmark of Gervais-Neveu
quantization and guarantees the  preservation
of the residual  symmetry Eq.\ref{2.6}. From
Eqs.\ref{2.4}, \ref{freefields} we have that the fields $f_m^{(J)}$
can be written as products of exponentials of the $\vartheta_1$
and $\vartheta_2$ fields:
\beq
f_m^{(J)}=e^{(J-m)\sqrt{\gamma}\vartheta_1}e^{(J+m)\sqrt{\gamma}
\vartheta_2}
\label{exprep}
\eeq
Though this form is, in principle,
accessible directly to quantization  for any $J$
and $m$ (cf. ref.\cite{GS1} ),  for the purposes of the present
analysis
it is more appropriate to work with an alternative Coulomb-gas-type
representation
in terms of one free field only. Let us consider the special cases
$m=\pm J$ of Eq.\ref{exprep} where we have
\beq
f^{(J)}_{-J}=({1\over \sqrt{A'}})^{2J}=
e^{ 2J\sqrt{\gamma}\vartheta_1}, \qquad
f^{(J)}_{J}=({A\over \sqrt{A'}})^{2J}=e^{
2J\sqrt{\gamma}\vartheta_2},
\label{2.10}
\eeq
Using the  periodicity requirement Eq.\ref{2.9},
one easily derives the relations
\beqa
A(u)&=&
\left \{ e^{-4\pi p_0^{(1)} \sqrt{\gamma} }
\int_0^uf^{(-1)}_{1}(\rho)d\rho +
\int_u^{2\pi}f^{(-1)}_{1}(\rho)d\rho \right \} \left /
\left ( e^{-4\pi p_0^{(1)} \sqrt{\gamma} } -1\right )\right. \quad
\label{2.11} \\
{-1\over A(u)}&=&
\left \{ e^{-4\pi p_0^{(2)} \sqrt{\gamma} }
\int_0^uf^{(-1)}_{-1}(\rho)d\rho +
\int_u^{2\pi}f^{(-1)}_{-1}(\rho)d\rho \right \} \left  /
\left ( e^{-4\pi p_0^{(2)} \sqrt{\gamma} } -1 \right)\right.  \quad
\label{2.12}
\eeqa
Then we may rewrite Eq.\ref{2.4} as
\beq
f_m^{(J)}= \sqrt{\scriptstyle{2J \choose J+m}} f_{-J}^{(J)}
A^{J+m} =\sqrt{\scriptstyle{2J \choose J+m}} f_{J}^{(J)}
(A^{-1})^{J-m}
\label{2.13}
\eeq
The starting point of the quantization is  to replace Eqs.\ref{2.8}
by their quantum counterparts, so that we now have
\begin{equation}
\Bigl [\vartheta'_1(\sigma_1),\vartheta'_1(\sigma_2) \Bigr ]=
\Bigl[\vartheta'_2(\sigma_1),\vartheta'_2(\sigma_2) \Bigr ]
=2\pi i\,  \delta'(\sigma_1-\sigma_2)
\label{2.14}
\end{equation}
It was shown in ref.\cite{GN3} that $\vartheta_1$ and $\vartheta_2$
are related by a complicated canonical transformation;
however, the relation between the zero modes is simple:
\beq
p_0^{(1)}=-p_0^{(2)} \quad ,\quad \bar p_0^{(1)}=-\bar p_0^{(2)}.
\label{zeromodes}
\eeq
Instead of $p_0^{(1)}$ (or $p_0^{(2)}$) it will be more convenient to
work with the rescaled zero mode
\beq
\varpi :=ip_0^{(1)}\sqrt{2\pi\over h}
\label{defomega}
\eeq
with $h$ defined in terms of the central charge $C$ by
\beq
h={\pi \over 12}(C-13 -\sqrt{(C-25)(C-1)})
\label{defh}
\eeq
The  parameter $h$ which is  the deformation parameter
of $sl(2)$, is also in effect the  Planck constant of the quantum
Liouville
theory.

\section{The case of a single screening charge}
\subsection{The braiding of the holomorphic components}
Starting from the representation
Eq.\ref{2.13}, and following the method of ref.\cite{GS1},  we
construct the
quantum equivalents of the fields $f_m^{(J)}$. In the original work,
they have been noted $V_m^{(J)}$, $\Vt_m^{(J)}$, or
$U_m^{(J)}$, depending upon the normalization chosen. They are
periodic
up to a multiplicative constant and thus can be considered as Bloch
waves.
On the other hand, there is also a second basis of chiral operators
$\xi_M^{(J)}$,
which are by construction explicitly covariant under the quantum
group
\cite{G1}\cite{CGR2}, and related to the Bloch wave fields by a
linear
transformation. In the present article, we will concentrate on the
Bloch wave basis, except for the last section. We first explain the
construction of the Bloch wave vertex operators and their exchange
algebra, in part as a preparation for the case of two screening
charges.
The presentation follows ref.\cite{GS3}.
As discussed above, we consider the semi-infinite families of Bloch
wave
operators with $J+m$ or $J-m$ a non-negative integer.
It turns out that there exists  a consistent operator algebra where
the two types of families
do not mix. Thus
  we may  concentrate on one type, say  the case
  with $J+m=0,1,\dots$.
Then  the quantum version
of $f^{(J)}_{m}$ is  most easily obtained
from the quantum versions of (the
left equality in) Eq.\ref{2.13}, and  of Eq.\ref{2.11}.
According to Eq.\ref{2.10}, this leads to quantum expressions in
terms
of $\vartheta_1 $ --- note that the other case ( $J-m$ integer)  may
be obtained by the replacement
$\vartheta_1 \leftrightarrow \vartheta_2$ everywhere (cf. also
section
\ref{theta1theta2inv}).

To begin with, the factor $f^{(J)}_{-J}$
is replaced by
 the normal-ordered exponential
\beq
f_{-J}^{(J)}\bigr |_{\hbox {\scriptsize qu}}
\equiv U_{-J}^{(J)}=
: e^{2J\sqrt{h/2\pi}\vartheta_1}:
\label{3.0}
\eeq
The parameter $h$ is that of Eq.\ref{defh}.
The change of the coefficient in the exponential is such that this
field
has conformal weight
\begin{equation}
\Delta_J=-J -{h\over \pi}J (J +1).
\label{2.16}
\end{equation}
If  $2J$ is a positive integer, this coincides with Kac's formula,
and $U_{-J}^{(J)}$ is a
$(1, 2J+1)$ primary in the BPZ classification.
 Now let us turn to the second factor $A^{J+m}$ appearing on the
left equation of
Eq.\ref{2.13}.
The classical expression Eq.\ref{2.11}  for $A$
 has a rather simple quantum generalization, which we
will denote by  $S$ to signify that it is
is a primary field of dimension zero (``screening operator''),
namely\cite{LuS}
\beq
 S(\sigma)= e^{2ih(\varpi+1)}
 \int _0 ^{\sigma} d\rho U_{1}^{(-1)}(\rho)+
 \int _\sigma  ^{2\pi } d\rho U_{1}^{(-1)}(\rho)
\label{3.1}
\eeq
Apart from an overall change of normalization ---
removal of the denominator ---  and the introduction of normal
orderings,
 the only change
 consists in the replacement
$\varpi \to \varpi+1$ in the prefactor of the first integral.
 The quantum formula is such that $S$ is
periodic up to a multiplicative factor
\beq
S(\sigma+2\pi)=e^{2ih(\varpi+1)}S(\sigma).
\label{3.2}
\eeq
This is the quantum version of the left equation in Eq.\ref{2.9}.
$S$ is the operator formed from the semiclassical screening charge
\beq
\alpha_-\equiv 2\sqrt{h\over 2\pi}
\label{screeningcharge}
\eeq
which tends to zero in the classical limit $h\to 0$, or $C\to\infty$
 (cf. Eq.\ref{defh}). $\alpha_-$ is just the same quantity that
appears
e.g. in the treatment of minimal models\cite{DF}, except that there
the
central charge is smaller than one.
The basic primary field of the Coulomb gas picture is now
defined as
\beq
U_{m}^{(J)}(\sigma)=  U_{-J}^{(J)}(\sigma) [S(\sigma)]^{J+m}
\label{3.3}
\end{equation}
which is the quantum version of the first equality in Eq.\ref{2.13}.
The product of operators at the same point implied in Eq.\ref{3.3}
exists for small enough $h$ (more on this  below).
Since $S$ is a screening operator, the conformal dimension of
$U_m^{(J)}$
 agrees
with Eq.\ref{2.16}. It should be emphasized that the Coulomb gas
picture
obtained here differs from the standard one\cite{DF} in that it is
"operatorial", i.e. there are no contours which depend on the
particular
correlation function under consideration.
It is easy
to verify that
\begin{equation}
U_m^{(J)} \varpi = (\varpi+2m) U_m^{(J)}
\label{3.4}
\end{equation}
Here we are assuming $\varpi$ to be real, as is appropriate in
the socalled
elliptic sector of the theory (cf. section \ref{liouexp}). Also
in the rest of the review we will concentrate on this case, if not
indicated otherwise. It is the case which appears to be directly
related to (tree level) amplitudes in $c\le 1$ string
theory\cite{G5}.
The normalization of the $U_m^{(J)}$ operators is given by
\beq
<\varpi |U^{(J)}_m |\varpi+2m>=I^{(J)}_m(\varpi)
\label{3.5}
\eeq
and $I^{(J)}_m(\varpi)$ can be computed to be \cite{GS3}
\footnote{see ref.\cite{Fe} for a
similar calculation, applied to the degenerate case}
\[
I_m^{(J)}(\varpi )=
\left( 2 \pi \Gamma(1+{h\over\pi}) \right )^{J+m}
e^{ ih(J+m)(\varpi-J+m)}
\]
\beq
\prod_{\ell=1}^{J+m}
{\Gamma[1+(2J-\ell+1)h/\pi] \over \Gamma[1+\ell h/\pi]
 \Gamma[1-(\varpi+2m-\ell)h/\pi]
\Gamma[1+(\varpi+\ell)h/\pi] }.
\label{3.6}
\eeq
This formula illustrates  an important
point to be made about the integral representation Eq.\ref{3.3}.
For small enough $h$, the arguments of the gamma functions are
all positive,
and this corresponds to the domain where the integral
representation is
convergent. When $h$ increases, divergences
 appear. However, Eq.\ref{3.6}
continues to make sense beyond the poles by the usual
analytic continuation
of the Gamma function. As is well known\cite{GN4}, the continuation
of the ground state expectation value $I_m^{(J)}(\varpi )$ defines
the
continuation of the operator $U_m^{(J)}$ itself. Thus, $U_m^{(J)}$
is related to  the normalized operator $V_m^{(J)}$ with
$\langle \varpi |V_m^{(J)}(\sigma =0) |\varpi +2m\rangle =1$
(introduced already in ref.\cite{CGR1} for half-integer positive $J$)
by
\beq
U^{(J)}_m =I^{(J)}_m(\varpi) V^{(J)}_m \quad .
\label{3.7}
\eeq
Note that $U_{-J}^{(J)}\equiv V_{-J}^{(J)}$.

We now come to
 the braiding  algebra of the fields
$U_m^{(J)}$. For half-integer positive spin --- the case
corresponding to Kac's table --- it is well known that
the braiding of the $U_m^{(J)}$ or $V_m^{(J)}$ is essentially given
by a q-$6j$-symbol, and the explicit formulae were
determined in ref.\cite{CGR1} (The general result is summarized in
appendix A).
In ref.\cite{GS1},  this result
was extended to arbitrary $J$.
We will recall some basic points of the derivation
 that will be useful later on.
The braiding relation takes the form
\beq
U_m^{(J)}(\sigma) U_m^{(J')}(\sigma')=
\sum_{m_1,m_2} R_U(J,J';\varpi)^{m_2 m_1}_{m\phantom{_2}
m'\phantom{_1}}
U_{m_2}^{(J')}(\sigma') U_{m_1}^{(J)}(\sigma)
 \label{3.10}.
\eeq
We only deal with the case $2\pi >\sigma '>\sigma >0$ explicitly. The
other
cases are deduced from the present one, by using the periodicity
properties of the $U_m^{(J)}$ fields, and the fact that the
$R$-matrix
for the other order of $\sigma$ and $\sigma'$ must be just the
inverse of the one in Eq.\ref{3.10}.
The sums extend over non-negative integer $J+m_1$ resp. $J'+m_2$
with the condition
\begin{equation}
m_1+m_2=m+m'=:m_{12}.
\label{3.11}
\end{equation}
Since one considers the braiding at
equal $\tau$ one can let $\tau=0$ once and for all.
As there are no  null-vector
decoupling equations for continuous $J$,
the derivation of Eq.\ref{3.10}   relies  exclusively on the free
field techniques summarized in the previous section.
The basic point of our argument
is that
 the exchange of two $U_m^{(J)}$
operators can be mapped into an equivalent problem in one-dimensional
quantum mechanics, and becomes just finite-dimensional linear
algebra.
In  view of Eqs.\ref{3.1}, \ref{3.3},
 the essential observation
is  that one only needs the braiding relations of
$U_{-J}^{(J)}\equiv V_{-J}^{(J)}$ operators which are normal ordered
exponentials
(``tachyon operators''). One has
\begin{equation}
V_{-J}^{(J)}(\sigma )V_{-J'}^{(J')}(\sigma ')=
e^{-i2JJ' h \epsilon (\sigma -\sigma')}
V_{-J'}^{(J')}(\sigma ' )V_{-J}^{(J)}(\sigma )
\label{3.12}
\end{equation}
where $\epsilon (\sigma -\sigma')$ is the sign of $\sigma -\sigma'$.
The fact that Eq.\ref{3.12} depends on the position only through
the order of $\sigma$ and $\sigma'$ is crucial for the following
argument; it means that the problem is in a sense of a topological
nature.
Indeed, the results we obtain for the single screening case
are in agreement with those obtained
by an analysis of the braiding problem for the Liouville theory on
the
lattice\cite{B2}, though the language is rather different, and
$6j$ symbols do not
explicitly appear in ref.\cite{B2}.
The first step of the analysis is to note that when commuting the
tachyon operators in
$U_{m'}^{(J')}(\sigma ')$ through those of
$U_{m}^{(J)}(\sigma )$, one only encounters phase factors of the form
$e^{\pm i2 \alpha \beta h } $ resp. $e^{\pm  6i \alpha \beta h } $,
with $\alpha$ equal to $J$ or $-1$, $\beta$ equal to $J'$ or $-1$,
since  we take
$\sigma, \sigma' \in  [0,2\pi]$. Hence we are led to decompose the
integrals defining the screening charges S into pieces which commute
with each other and  with $V_{-J}^{(J)}(\sigma)$, $V_{-J'}^{(J')}
(\sigma ')$ up to one of the above phase factors. We consider
explicitly only the case $0< \sigma < \sigma ' < 2\pi$ and write
\[
S(\sigma )\phantom{'} = S_{\sigma \sigma '} + S_\Delta,
\quad
S(\sigma ')= S_{\sigma \sigma '} + k(\varpi)S_\Delta \equiv
S_{\sigma \sigma '} + \tilde {S}_\Delta,
\]
\[
S_{\sigma \sigma '}:= k(\varpi)
\int_0^\sigma V_{1}^{(-1)}(\rho )
d\rho + \int_{\sigma '}^{2\pi}V_{1}^{(-1)}(\rho )
d\rho,
\]
\begin{equation}
S_\Delta := \int_\sigma^{\sigma '}V_{1}^{(-1)}(\rho )
d\rho,
\quad
k(\varpi):= e^{2ih(\varpi+1)}
\label{3.13}
\end{equation}
Using Eq.\ref{3.12}, we then get the following simple algebra for
$S_{\sigma \sigma '}, S_\Delta , \tilde {S}_\Delta$:
\medskip
\begin{equation}
S_{\sigma \sigma '}S_\Delta =q^{-2}S_\Delta S_{\sigma \sigma '},
\quad
S_{\sigma \sigma '}\tilde {S}_\Delta =q^2 \tilde {S}_\Delta
S_{\sigma \sigma '},
\quad
S_\Delta \tilde {S}_\Delta =q^4 \tilde {S}_\Delta S_\Delta,
\label{3.14}
\end{equation}
and their commutation properties with
$V_{-J}^{(J)}(\sigma ),V_{-J'}^{(J')}(\sigma ' )$ are given by
 \begin{equation}
\begin{array}{lll}
&V_{-J}^{(J)}(\sigma) S_{\sigma \sigma '}=q^{-2J}S_{\sigma \sigma'}
V_{-J}^{(J)}(\sigma ),
&V_{-J'}^{(J')}(\sigma')S_{\sigma \sigma '}= q^{-2J'} S_{\sigma
\sigma'}
V_{-J'}^{(J')}(\sigma'),
\nonumber\\
&V_{-J}^{(J)}(\sigma)S_\Delta  = q^{-2J}S_\Delta V_{-J}^{(J)}(\sigma
),
&V_{-J}^{(J)}(\sigma) \tilde {S}_\Delta \  =q^{-6J} \tilde {S}_\Delta
V_{-J}^{(J)}(\sigma),
\nonumber \\
&V_{-J'}^{(J')}(\sigma')S_\Delta = q^{2J'} S_\Delta V_{-J'}^{(J')}
(\sigma '),
&V_{-J'}^{(J')}(\sigma')\tilde {S}_\Delta = q^{-2J'} \tilde
{S}_\Delta
V_{-J'}^{(J')}(\sigma').
\end{array}
\label{3.15}
\end{equation}
Finally, all three screening pieces obviously shift the zero mode in
the
same way:
\begin{equation}
 \left. \begin{array}{ccc}
S_{\sigma \sigma '}\nonumber \\ S_\Delta \nonumber \\ \tilde
{S}_\Delta
\end{array} \right \} \varpi
= (\varpi +2) \left \{ \begin{array}{ccc}
S_{\sigma \sigma '}\nonumber \\ S_\Delta \nonumber \\ \tilde
{S}_\Delta
\end{array} \right. .
\label{3.16}
\end{equation}
Using Eqs.\ref{3.15} we can commute $\Vt_{-J}^{(J)}(\sigma  )$ and
$\Vt_{-J'}^{(J')}(\sigma  ')$ to the left on both sides of
Eq.\ref{3.10},
so that they can be cancelled. Then we are left with
\[
(q^{-2J'}S_\Delta + q^{2J'}S_{\sigma \sigma '} )^{J+m}
(\tilde {S}_\Delta +S_{\sigma \sigma '})^{J'+m'}q^{2JJ'}=
\]
\begin{equation}
\sum_{m_1,m_2} R(J,J';\varpi + 2(J+J'))_{m_{\phantom{2}}m'}^{m_2 m_1}
(q^{2J} S_{\sigma \sigma '} + q^{6J}\tilde {S}_\Delta )^{J'+m_2}
(S_{\sigma \sigma '} + S_\Delta )^{J+m_1}
\label{3.17}
\end{equation}
It is apparent from this equation that the braiding problem of the
$U_m^{(J)}$ operators is governed by the Heisenberg-like algebra
Eq.\ref{3.14}, characteristic of one-dimensional
quantum mechanics. We will  proceed using  the following simple
representation
of the algebra Eq.\ref{3.14} in terms of one-dimensional
quantum mechanics ( $y$ and $y'$ are arbitrary complex numbers):
\begin{equation}
S_{\sigma \sigma '}=y' e^{2Q}, \quad
S_\Delta =y e^{2Q-P}, \quad  \tilde {S}_\Delta =
y e^{2Q+P},
 \qquad [Q,P]=ih  .
\label{3.18}
\end{equation}
The third  relation in Eq.\ref{3.18} follows from
the second one in view of
$\tilde {S}_\Delta =k(\varpi )S_\Delta$ (cf. Eq.\ref{3.13}).
        This means we are
identifying here $P \equiv ih\varpi $ with the zero mode of the
original
problem. Using $e^{2Q+cP} = e^{cP} e^{2Q} q^c$ we can commute all
factors
$e^{2Q}$ to the right on both sides of Eq.\ref{3.17} and then cancel
them.
This leaves us with
\[
q^{2JJ'}
\prod_{s=1}^{J+m}(y'q^{2J'} +yq^{-(\varpi -2J+2s-1)})
\prod_{t=1}^{J'+m'} (y'+yq^{\varpi -2J'+2m+2t-1})=
\]
\begin{equation}
\sum_{m_1} R_U(J,J';\varpi)_{m_{\phantom {2}} m'}^{m_2 m_1}
\prod_{t=1}^{J'+m_2}(y'q^{2J} +yq^{\varpi +4J-2J'+2t-1})
\prod_{s=1}^{J+m_1} (y'+yq^{-(\varpi -2J+2m_2 +2s -1)})
\label{3.19}
\end{equation}
where we have shifted back $\varpi +2(J+J') \rightarrow \varpi$
compared to Eq.\ref{3.17}.
Since the overall scaling $ y\rightarrow \lambda y,
y' \rightarrow \lambda y'$ only gives back Eq.\ref{3.11}, we can set
$y'=1$.

The solution of these equations, which was
 derived in ref.\cite{GS1}, will be
cast under the convenient form
\beq
R_U(J,J',\varpi)_{m_{\phantom{2}} m'}^{m_2 m_1}=
e^{-i\pi (\Delta_c+\Delta_b
-\Delta_e-\Delta_f)}
{\kappa_{ab}^e \kappa_{de}^c\over \kappa_{db}^f \kappa_{af}^c}
\left\{ ^{a}_{d}\,^{b}_{c}
\right. \left |^{e}_{f}\right\}
\label{3.20}
\eeq
where $\left\{ ^{a}_{d}\,^{b}_{c}
\right. \left |^{e}_{f}\right\}$ is the q-$6j$-symbol generalized
to continous spins, with arguments
$$
a=J,\quad b=x+m+m',\quad c=x\equiv (\varpi-\varpi_0)/2
$$
\beq
d=J',\quad e=x+m_2,\quad f=x+m,
\label{3.21}
\eeq
and the coefficients $\kappa_{J_1 J_2}^{J_{12}}$ are given by
\footnote{We follow the prescription of refs.\cite{G5},\cite{GR} for
the definition of the square roots (cf. also below Eq.\ref{defg}).
In particular, the square root in Eq.\ref{3.9} is to be understood
as the product resp. quotient of the square roots of the individual
factors consisting of a single q-number.}
$$
\kappa_{J_1 J_2}^{J_{12}}=
\left ({ he^{-i(h+\pi)}
\over 2\pi \Gamma(1+h/\pi) \sin h}\right )^{J_1+J_2-J_{12}}
e^{ih (J_1+J_2-J_{12})( J_1-J_2-J_{12})} \times
$$
\beq
\prod_{k=1}^{J_1+J_2-J_{12}}
\sqrt{ \lfloor 1+2J_1-k\rfloor \over
\lfloor k \rfloor \, \lfloor 1+2J_2-k \rfloor\,
\lfloor -(1+2J_{12}+k) \rfloor }.
\label{3.9}
\eeq
Recall that we let $\lfloor x \rfloor =\sin(hx)/\sin h$ in general.
The  last equation  makes sense
for arbitrary $J_1,J_2, J_{12}$ such that $J_1+J_2-J_{12}$ is a
non-negative integer.
The r.h.s.of Eq.\ref{3.20}
 may be expressed in terms of q-hypergeometric functions
by the formula
\[
{\kappa_{ab}^e \kappa_{de}^c\over \kappa_{db}^f \kappa_{af}^c}
\left\{ ^{a}_{d}\,^{b}_{c}
\right. \left |^{e}_{f}\right\}
=
q^{2dn-2an_2}\lfloor 2e+1\rfloor
{\lfloor \phi \rfloor_{n_1}
 \lfloor 1-\epsilon-n_1\rfloor_{n_1} \over
\lfloor n_1 \rfloor \! !
\lfloor \beta-\epsilon-n'-n \rfloor_{n+n'+1}
 }\times
\]
\beq
\times\lfloor \rho\rfloor_{n'}
\lfloor 1-\beta\rfloor_{n-n_1}
\>_4 F_3\left (^{\alpha,}_{\epsilon,} \,\,
^{\beta ,}_{\phi ,} \,\, ^{-n_1 ,}_
{\rho } \,\, ^{-n'} \ ;q,1 \right )
\label{3.22}
\eeq
with\hfill
$$
\alpha=-a-c+f,\quad \beta=-c-d+e,
$$
$$
n_1=a+b-e, \quad n_2=d+e-c,\quad  n=f+a-c \quad n'=b+d-f;
$$
\beq
\epsilon=-(a+b+c+d+1), \quad
\phi=1+n-n_1, \quad
\rho=e+f-a-d+1.
\label{3.23}
\eeq
Here we have  defined
\[
_4 F_3 \left (^{a,}_{e,} \, ^{b,}_{f,} \,
^{c,}_{g} \, ^{d} ;q,\rho \right )=
\sum_{n=0}^\infty {\lfloor a\rfloor_n \lfloor b\rfloor_n
\lfloor c\rfloor_n \lfloor d\rfloor_n \over \lfloor e\rfloor_n
\lfloor f\rfloor_n \lfloor g\rfloor_n \lfloor n\rfloor !} \ \rho^n ,
\]
\beq
\lfloor a\rfloor_n := \lfloor a\rfloor \lfloor a+1\rfloor \cdots
\lfloor a+n-1\rfloor ,
\quad \lfloor a\rfloor_0 :=1.
\label{3.25}
\eeq
 In  Eq.\ref{3.22}, the prefactor involves
 products of the type just recalled with
indices
\beq
n_1=J+m_1, \quad
n_2=J+m_2, \quad
n=J+m, \quad
n'=J+m'.
\label{3.24}
\eeq
Since  they are equal to the screening numbers,
they  are positive integers. Thus  Eq.\ref{3.22}
makes sense for arbitrary spins provided the screening numbers are
integers, and is the appropriate generalization.

The method used to derive Eq.\ref{3.20} was to
transform\footnote{This
transformation formula is derived in ref.\cite{GR}.} the
hypergeometric function into another one such that
the desired relations
Eq.\ref{3.19}
follow from the orthogonality relation of the associated Askey-Wilson
polynomials. In this connection, let us  note that a
simple reshuffling of the parameters of the latter form
 allows to verify that
the usual orthogonality relations of the 6-j symbols extend
to our case. One has,  in general,
\beq
\sum_{J_{23}}
\left\{ ^{J_1 }_{J_3 }
\> ^{J_2 } _{ J_{123}}
\right.
\left |^{ J_{12}}
_{J_{23}}\right\}
\left\{ ^{J_1 }_{J_3 }
\> ^{J_2 } _{ J_{123}}
\right.
\left |^{ K_{12}}
_{J_{23}}\right\} =\delta_{J_{12}-K_{12}}
\label{orth}
\eeq
where the $J$'s are arbitrary except for the constraint that
  the screening numbers
$$
n_1=J_1+J_2-J_{12}, \quad n_2=J_3+J_{12}-J_{123}, \quad
n=J_1+J_{23}-J_{123},\quad n'=J_2+J_3-J_{23},
$$
\beq
\tilde n_1=J_1+J_2-K_{12}, \quad \tilde n_2=J_3+K_{12}-J_{123},
\label{3.26}
\eeq
are positive integers. These  conditions fix the range of summation
over
$J_{23}$.

The basis $U_m^{(J)}$, apart from its manageability, has another
practical
virtue: Its braiding (and fusion) properties  are given in a form
which involves no square roots, but
only (q-deformed) rational functions, and no phase
ambiguities can arise. On the other hand,
from the quantum group point of view
it is more natural to consider a basis where
the braiding (and fusion) is given exclusively
in terms of the $6j$-symbol (the latter does however
involve square roots).
For this purpose, the authors of ref.\cite{CGR1} introduced the
fields
$\Vt_m^{(J)}$, which are defined by
\beq
\Vt_m^{(J)}=g_{J,x+m}^{x} V_m^{(J)}.
\label{3.27}
\eeq
As before we let  $x=(\varpi-\varpi_0)/2$. The coupling constants
$g$ are defined by
$$
g_{J, \, x +m}^{x}=
\left ( {h\over\pi}\right)
^ {J+m}
\prod_{k=1}^{J+m}
\sqrt{
F[1+(2J-k+1)h/\pi]}
\sqrt{F[(\varpi+2m-k)h/\pi]} \times
$$
\beq
\prod_{k=1}^{J+m} \sqrt{ F[-(\varpi+k)h/\pi]} \left  /
\sqrt{F[1+kh/\pi]}\right.
\label{defg}
\eeq
where, as usual,  $F(z):=\Gamma(z)/\Gamma(1-z)$. The treatment of the
square roots requires some care. We follow the prescription
of ref.\cite{G5}
also used in ref.\cite{GR}. Eq.\ref{defg} immediately extends to
the case of non integer $J$, as $J+m$ remains a positive integer.
We then have ($m_{12}:=m_1+m_2=m+m'$)
\[
\Vt_m^{(J)}(\sigma)\Vt_{m'}^{(J')}(\sigma')=
e^{-i\pi(\Delta_{x} +\Delta_{x+m_{12}}
-\Delta_{x+m} -\Delta_{x+m_2})}\times
\]
\beq
\sum_{m_1, m_2}
\left\{ ^{J\quad }_{J'\quad }
\> ^{x +m_{12}}
_{ x\phantom{+m_{12}}}
\right.
\left |^{\quad  x+m_2}
_{\quad  x+m}\right\}
\Vt_{m_2}^{(J')}(\sigma')\Vt_{m_1}^{(J)}(\sigma)
\label{3.29}
\eeq
The relation with $U^{(J)}_m$ is given by
\beq
U^{(J)}_m ={I^{(J)}_m(\varpi)\over
g_{J, \, x +m}^{x}} \Vt^{(J)}_m
\equiv {1\over \kappa_{J, \, x +m}^{x}}
\Vt^{(J)}_m.
\label{3.8}
\eeq
The $\kappa$ coefficient is of course given by Eq.\ref{3.9}.
Concerning the right-moving
modes, the braiding algebra is given by
\[
\Vtb_m^{(J)}(\sigma)\Vtb_{m'}^{(J')}(\sigma')=
e^{i\pi(\Delta_{\xb} +\Delta_{\xb+m_{12}}
-\Delta_{ \xb+m} -\Delta_{\xb+m_2})}\times
\]
\beq
\sum_{m_1, m_2}
\left\{ ^{J_1\quad }_{J_2\quad }
\> ^{\xb +m_{12}}
_{ \xb\phantom{+m_{12}}}
\right.
\left |^{\quad  \xb+m_2}
_{\quad  \xb+m}\right\}
\Vtb_{m_2}^{(J')}(\sigma')\Vtb_{m_1}^{(J)}(\sigma)
\label{Vtbar}
\eeq
where we let $\xb=(\varpib-\varpi_0)/2$.
The only difference with Eq.\ref{3.29} is the change of sign of the
phase factor. This may be verified  by redoing the whole derivation.
In refs.\cite{LuS},\cite{G5} it was remarked that the right-mover
braiding matrix
is
deduced from the left-mover one by changing $i=\sqrt{-1}$ into $-i$,
since
this correctly changes the orientation of the complex plane.
This complex conjugation is most easily performed using the $U$
fields,
since the braiding matrix Eq.\ref{3.20} is real apart from the
first phase factor. For the $\Vt$ fields,
there is a
slight subtlety
 related again
to the appearance of the redundant
square roots in Eq.\ref{3.9}. The  correct rule is to
take the same definition for the square roots for left and right
movers.
Thus  the right-moving
coupling constant
$\bar g^{\xb}_
{J,\xb+m}$
is  given
by the {\it same} expression Eq.\ref{defg}, not its complex
conjugate.
The same
prescription should be followed for the roots appearing in
${\overline \kappa}^{\xb}_{J,\xb+m}$,
while taking the usual complex conjugate for the phase factor
appearing in front
of the product in Eq.\ref{3.9}. Note that $\varpi$ is always to be
        treated as
real formally in this context, even in the hyperbolic sector where it
is
actually purely imaginary (cf. below).

\section{Solving of the Liouville quantum dynamics}
\subsection{The Liouville exponential}
\label{liouexp}
First, let us note   that, as  $h$ is real
 in the weak-coupling regime,
 the hermiticity of energy-momentum
allows for $\varpi,\varpib$ real or purely imaginary, corresponding
to
the elliptic resp. hyperbolic sector of the theory\cite{JKM}
(see also \cite{CG2} for the case of open boundary conditions).
In the former case, which we consider in this review, we will see
that
the locality conditions are fulfilled if
\begin{equation}
\varpi-\varpib=k\pi/h,\qquad\qquad k\in {\bf Z}
\label{condx}
\end{equation}
Eq.\ref{condx} has an immediate interpretation as the natural
generalization
of the classical boundary conditions\cite{LuS}.
Moreover we will show that the appropriate definition of the
Liouville exponential  for arbitrary $J$ is
\beq
e^{\textstyle -J\alpha_-\Phi(\sigma, \tau )}=
\sum _{m=-J}^{\infty} \mu_0^{J+m}
\Vt_m^{(J)}(u)\,
{\overline \Vt_{m}^{(J)}}(v) \>
\label{3.31}
\eeq
where $\alpha_- =2\sqrt{h/2\pi}$ is the screening charge.
The constant $\mu_0^{J+m}$ will not be fixed by braiding or fusion.
It will be determined below when we derive  the field equations.
\subsubsection{Locality}
Let us now check locality.
In the approach of refs.\cite{LuS}\cite{G5}, one takes
the zero modes of the left-moving and right-moving Liouville modes
to commute, so that $U$ and ${\overline U}$ commute. However,
operators
involving both chiralities should be applied only to states
fulfilling Eq.\ref{condx}, and conserve this condition. This is why
we
must
have
$\bar m=m$ in Eq.\ref{3.31}.\footnote{The situation is quite
different
in the strong coupling theory, see e.g. ref.\cite{GR}.}
Next we observe that if Eq.\ref{condx}
is valid,
 \beq
{\overline R}_{\Vtb}(J,J';{\overline\varpi})
^{\bar m_2 \bar m_1}_{\bar m_{\phantom{2}} \bar m'}=
{\overline R}_{\Vtb}(J,J';\varpi)^{\bar m_2 \bar m_1}_{\bar
m_{\phantom{2}}
\bar m'}
\label{omegabar}
\eeq
as can be verified easily. The same is  true for ${\overline
R}_{\overline U}$.
 Thus we have
\[
\Vtb_m^{(J)}(\sigma)\Vtb_{m'}^{(J')}(\sigma')=
e^{+i\pi(\Delta_{x} +\Delta_{x+m_{12}}
-\Delta_{x+m} -\Delta_{x+m_2})}\times
\]
\beq
\sum_{m_1, m_2}
\left\{ ^{J_1\quad }_{J_2\quad }
\> ^{x+m_{12}}
_{ \quad \quad x}
\right.
\left |^{\quad  x +m_2}
_{\quad  x+m}\right\}
\Vtb_{m_2}^{(J')}(\sigma')\Vtb_{m_1}^{(J)}(\sigma)
\label{Vtbar2}
\eeq
Then,
according to Eqs.\ref{3.29}, \ref{Vtbar2} we get
$$
e^{\textstyle -J\alpha_-\Phi(\sigma, \tau )}
e^{\textstyle -J'\alpha_-\Phi(\sigma', \tau )}=
$$
$$
\sum_{m, m'}
\sum_{m_1, m_2; \mb_1, \mb_2}e^{ih(\mb_2-m_2)(\mb_2-m_2 +\varpi)}
\left\{ ^{J\quad }_{J'\quad }
\> ^{x +m_1+m_2}
_{x\phantom{+m_1+m_2}}
\right.
\left |^{ x+m_2}
_{ x+m}\right\}
\left\{ ^{J\quad }_{J'\quad }
\> ^{x+m_1+m_2}
_{x\phantom{+m_1+m_2}}
\right.
\left |^{x +\mb_2}
_{x+m}\right\}
$$
\beq
\times  \mu_0^{J+m+J'+m'}
\left. \Vt_{m_2}^{(J')}(u') \Vt_{m_1}^{(J)}(u)\,
{\overline \Vt_{m_2}^{(J')}}(v') {\overline \Vt_{m_1}^{(J)}}(v)
 \right |_{m_1+m_2=m+m' \atop
{\overline m}_1+{\overline m}_2=m+m'},
\label{braidexp}
\eeq
  One first sums  over $m$, with fixed
$m_{12}=m+m'$. This precisely corresponds to the
 summation over $J_{23}$ in Eq.\ref{3.26}.
Thus only $m_2=\mb_2$ contributes.
This gives immediately
\beq
e^{\textstyle -J_1\alpha_-\Phi(\sigma_1, \tau )}
e^{\textstyle -J_2\alpha_-\Phi(\sigma_2, \tau )}=
e^{\textstyle -J_2\alpha_-\Phi(\sigma_2, \tau )}
e^{\textstyle -J_1\alpha_-\Phi(\sigma_1, \tau )},
\label{loc}
\eeq
and the Liouville exponential is local for arbitrary $J$.
We remark that Eq.\ref{condx} is not only sufficient, but also
necessary for locality, as was observed in ref.\cite{LuS} for the
special case $J=1/2$.

\subsubsection{Closure by fusion}
In the preceding analysis, we have discussed only the braiding
properties
of the chiral fields resp. the Liouville exponentials. However,
according
to the general Moore-Seiberg formalism\cite{MS}, fusion (in the sense
of the
full operator product) and braiding are not independent. Assuming
the validity of the Moore-Seiberg relation between fusion and
braiding
matrix,
we then obtain immediately
that the fusion of the
 $\Vt$ fields should be given by (cf. also ref.\cite{GR})
$$
\Vt^{(J_1)}_{m_1}(z_1) \Vt^{(J_2)}_{m_2}(z_2) =
\sum _{J_{12}= -m_1-m_2} ^{J_1+J_2}
\left\{ ^{ J_1}_{x +m_1+m_2}\,
\> ^{ J_2}_{x}
\right.
\left |^{ J_{12}}_{x+m_1}\right\}
\times
$$
\beq
\sum _{\{\nu_{12}\}}
\Vt ^{(J_{12},\{\nu_{12}\})}_{m_1+m_2}(z_2)
<\!\varpi_{J_{12}},{\{\nu_{12}\}} \vert
\Vt ^{(J_1)}_{J_2-J_{12}}(z_1-z_2) \vert \varpi_{J_2} \! >.
\label{fus1}
\eeq
In Eq.\ref{fus1} we have changed variables by letting
$z=e^{i(\tau+\sigma)}$, $\zb=e^{i(\tau-\sigma)}$ (recall that we are
using
Minkowski world-sheet variables). The only difference to the positive
half-integer spin case, which was completely analyzed in
ref.\cite{CGR1},
is that
the $J_{12}$ -sum now extends to $-m_1-m_2$ instead of $|J_1-J_2|$.
Indeed, the positivity of the screening numbers appearing in the
braiding
matrix leads via the Moore-Seiberg relation to the positivity of the
screening numbers $n_1=J_1+m_1,\  n_2=J_2+m_2,
\ p_{1,2}=J_1+J_2-J_{12},
 \ n=J_{12}+m_1+m_2$ of the fusion matrix. In
ref.\cite{GR}, it has been shown that the generalized $6j$-symbol
of Eq.\ref{3.20}, together with the positivity condition for the
screening charges, fulfills all the necessary identities for the
polynomial equations to be valid with continous spins. This provides
a
strong
argument that the fusion matrix of Eq.\ref{fus1} is indeed the
correct
one,
even though we have not attempted to derive it directly as we did for
the
braiding.
Making use of the analogous equation for the bar components, one sees
that the
operator-product expansion of Liouville exponentials may be written
as
$$
e^{\textstyle -J_1\alpha_-\Phi(z_1, \zb_1 )}
e^{\textstyle -J_2\alpha_-\Phi(z_2, \zb_2 )}=
$$
$$
\sum_{m_1,m_2} \sum _{J_{12}, \Jb_{12}}
\left\{ ^{ J_1}_{x +m_1+m_2}\,
\> ^{ J_2}_{x}
\right.
\left |^{ J_{12}}_{x+m_1}\right\}
\left\{ ^{ J_1}_{\xb +m_1+m_2}\,
\> ^{ J_2}_{\xb}
\right.
\left |^{ \Jb_{12}}_{\xb+m_1}\right\}\times
$$
$$
\mu_0^{J_1+m_1+J_2+m_2}
\sum _{\{\nu_{12}\}, \{\nub_{12}\} }
\Vt ^{(J_{12},\{\nu_{12}\})}_{m_1+m_2}(z_2)
\Vtb ^{(\Jb_{12},\{\nub_{12}\})}_{m_1+m_2}(\zb_2)\times
$$
\beq
<\!\varpi_{J_{12}},{\{\nu_{12}\}} \vert
\Vt ^{(J_1)}_{J_2-J_{12}}(z_1-z_2) \vert \varpi_{J_2} \! >
<\!\varpib_{\Jb_{12}},{\{\nub_{12}\}} \vert
\Vtb ^{(J_1)}_{J_2-\Jb_{12}}(\zb_1-\zb_2) \vert \varpib_{J_2} \! >
\label{fus2}
\eeq
It follows from condition Eq.\ref{condx}  that
$\left\{ ^{ J_1}_{\xb +m_1+m_2}\,
\> ^{ J_2}_{\xb}
\right.
\left |^{ \Jb_{12}}_{\xb+m_1}\right\}=
\left\{ ^{ J_1}_{x +m_1+m_2}\,
\> ^{ J_2}_{x}
\right.
\left |^{ \Jb_{12}}_{x+m_1}\right\}$. In the same way as for
locality,
the summation over $m_1$ with fixed $m_1+m_2$ then
reduces to the orthogonality relation for
$6j$-symbols so that only $J_{12}=\Jb_{12}$ contributes, and one gets
$$
e^{\textstyle -J_1\alpha_-\Phi(z_1, \zb_1 )}
e^{\textstyle -J_2\alpha_-\Phi(z_2, \zb_2 )}=
$$
$$
\sum_{m, J_{12}} \sum _{\{\nu_{12}\}, \{\nub_{12}\} }
\mu_0^{J_{12}+m}
\Vt ^{(J_{12},\{\nu_{12}\})}_{m}(z_2)
\Vtb ^{(J_{12},\{\nub_{12}\})}_{m}(\zb_2)
\mu_0^{J_1+J_2-J_{12}}\times
$$
\beq
<\!\varpi_{J_{12}},{\{\nu_{12}\}} \vert
\Vt ^{(J_1)}_{J_2-J_{12}}(z_1-z_2) \vert \varpi_{J_2} \! >
<\!\varpib_{J_{12}},{\{\nub_{12}\}} \vert
\Vtb ^{(J_1)}_{J_2-J_{12}}(\zb_1-\zb_2) \vert \varpib_{J_2} \! >.
\label{fus3}
\eeq
The second line clearly involves the descendants of the
Liouville exponentials
which we denote by
\beq
e^{\textstyle -J\alpha_-\Phi^{\{\nu\}, \{\nub\}}(z, \zb )}
\equiv
\sum_{m} \mu_0^{J+m}
\Vt ^{(J_{12},\{\nu\})}_{m}(z)
\Vtb ^{(J_{12},\{\nub\})}_{m}(\zb)
\label{fus4}
\eeq
As regards the last line, it is simply the corresponding matrix
element
of the
Liouville exponential. One finally gets
$$
e^{\textstyle -J_1\alpha_-\Phi(z_1, \zb_1 )}
e^{\textstyle -J_2\alpha_-\Phi(z_2, \zb_2 )}=
\sum_{ J_{12}=-m_1-m_2}^{J_1+J_2}  \sum _{\{\nu\}, \{\nub\} }
e^{\textstyle -J_{12} \alpha_-\Phi^{\{\nu\}, \{\nub\}}(z_2, \zb_2
)}\times
$$
\beq
< \!\varpi_{J_{12}}, \varpib_{J_{12}}; \{\nu\}, \{\nub\} |
e^{\textstyle -J_1\alpha_-\Phi(z_1-z_2, \zb_1-\zb_2 )} |
 \varpi_{J_2}, \varpib_{J_2} \! >.
\label{fus5}
\eeq
The notation for the matrix element should be self-explanatory
\footnote{It is implied here that charge conservation should be used
for
the evaluation of the matrix element, such that only the term
appearing
in
Eq.\ref{fus3} survives. According to ref.\cite{Cargese}, charge
conservation
actually does not hold for the 3-point functions
$\langle  \varpi'|e^{-J\alpha\Phi}(z)
|\varpi\rangle$ with continous $J$. From this point of view, the
notation of
Eq.\ref{fus5}
is of course not rigorously appropriate.}.
One sees that the Liouville exponential is closed by fusion for
arbitary
$J$ to all orders in the descendants.

\subsubsection{The cosmological constant revisited.}
The braiding relation is invariant under the transformation
\beq
\exp(-J\alpha_-\Phi(z, \zb )) \to T \exp(-J\alpha_-\Phi(z, \zb ))
T^{-1},
\label{simtrafo}
\eeq
where $T$ is an arbitrary function of the zero modes $\varpi$ and
$\varpib$.
This is
why locality does not completely determine the Liouville
exponentials.
We have discussed this point in detail in ref.\cite{GS2}.
Concerning  the fusion equation, the transformation just considered
does not act on the last term on the right-hand side which  is a
c-number.
The definition Eq.\ref{3.31} we have chosen is such that this term
 --- a  compact book-keeping device to handle all the descendants ---
is precisely given by the matrix elements of the Liouville
exponential
itself, without any additional normalization factor. It is thus
quite natural. The only remaining ambiguity  is the arbitrariness in
$\mu_0$.
Changing this parameter is  tantamount
to changing  the cosmological constant following ref.\cite{G5}.
Indeed,
the fusing and
braiding relations of the $\Vt$ fields are invariant if we  make the
change $\Vt_m^{(J)}\to \mu_c^{(J+m)/2} \Vt_m^{(J)}$.
Any such change is generated by a combination of a field
redefinition ($\alpha_-\Phi\to\alpha_-\Phi-\ln \mu_c$) and a
similarity
transformation of the form Eq.\ref{simtrafo}.
Thus the most general field satisfying Eqs.\ref{loc}
and \ref{fus5} is given by
\beq
e^{\textstyle -J\alpha_-\Phi(\sigma, \tau )}_{\mu_c}
=\sum _{m=-J}^{\infty}
(\mu_0 \mu_c)^{J+m}\Vt_m^{(J)}(u)\,
{\overline \Vt_{m}^{(J)}}(v)=\mu_c^{J} \mu_c^{-\varpi/2}
e^{\textstyle -J\alpha_-\Phi(\sigma, \tau )}\mu_c^{\varpi/2}
\label{muc}
\eeq
We will determine $\mu_0$ below so that it corresponds  to a
cosmological constant equal to one.

\subsubsection{Expression in terms of Coulomb-gas fields}

According to
Eqs.\ref{3.3} and \ref{3.8}, Eq.\ref{3.31} may be rewritten as
$$
e^{\textstyle -J\alpha_-\Phi(\sigma, \tau )}=
\sum _{m=-J}^{\infty}
\mu_0^{J+m} \kappa_{J, \, x +m}^{x}
\kappab _{J, \, \xb +m}^{\xb}
U^{(J)}_{m}(u) {\overline U}^{(J)}_{m}(v)
$$
It is easy to see using condition Eq.\ref{condx} that
\beq
\kappab_{J, \, \xb +m}^{\xb}= \kappab_{J, \, x +m}^{x}
\label{kappashift}
\eeq
Thus  the square roots combine pairwise
and we are left with a  rational expression.
$$
e^{\textstyle -J\alpha_-\Phi(\sigma, \tau )}=
\sum _{m=-J}^{\infty} \tilde \mu_0^{J+m} (-1)^{J+m}
\prod_{k=1}^{J+m}
{ \lfloor 1+2J-k\rfloor \over
\lfloor k \rfloor \, \lfloor \varpi+2m-k \rfloor\,
\lfloor \varpi +k \rfloor } \times
$$
\beq
V^{(J)}_{-J}(u) {\overline V}^{(J)}_{-J}(v) S^{J+m} \Sb^{J+m},
\label{Cbgas}
\eeq
where we have let
\beq
\tilde \mu_0= \mu_0 \left ({   h
\over 2\pi \Gamma(1+h/\pi) \sin h} \right )^2
\label{mu0tilde}
\eeq
Using this Coulomb-gas  expression,  together
with the mentioned orthogonality relations for Askey-Wilson
polynomials,
it is then possible to  directly verify  the locality of the
Liouville
exponential, without encountering any square root ambiguity.
Note that Eq.\ref{Cbgas} depends only on $\varpi$, not on $\varpib$.
On the other hand, the analysis of ref.\cite{LuS} for $J=1/2$ in the
elliptic sector, when translated to the Coulomb gas basis, gives
coefficients with an explicit dependence on $k$ of Eq.\ref{condx}.
Nevertheless, the two forms are equivalent, as they must, by means of
a
basis transformation Eq.\ref{simtrafo}, hence indistinguishable from
the point of view of locality.
\subsubsection{ $\vartheta_1 \leftrightarrow \vartheta_2$
invariance}
\label{theta1theta2inv}
In section \ref{self-contained} we noted the existence of a symmetry
of the theory under the exchange of the two free fields $\vartheta_1$
under $\vartheta_2$, the residual symmetry remaining after fixing
the $SL_2({\bf C})$ invariance. On the other hand, on the quantum
level
the expressions we have derived in the present review for the
Liouville
field and its exponentials are not evidently symmetric under
this exchange.
However, we must remember here that the requirement of locality
really
fixed these operators only up to a similarity transformation
Eq.\ref{simtrafo}
(the particular form Eq.\ref{3.31} resp. Eq.\ref{Cbgas} was only
distinguished
by its simplicity and its natural behaviour under fusion). Thus
a priori we can expect $\vartheta_1\leftrightarrow\vartheta_2$
invariance
only to be valid up to a similarity transformation. As a matter of
fact,
we will show  (for $J$ half-integer positive) that there exists
a transformation $T(\varpi,\varpib)$ such that
\beq
T(\varpi,\varpib) e^{-J\alpha_-\Phi}_{(1)} T^{-1}(\varpi,\varpib)=
T(-\varpi,-\varpib) e^{-J\alpha_-\Phi}_{(2)}
T^{-1}(-\varpi,-\varpib).
\label{sl2inv}
\eeq
where the index (1) resp. (2) indicates the
 use of the $\vartheta_1$ resp. $\vartheta_2$ representation.
This  shows in addition that
 it is possible to choose
 particular representatives in the equivalence class of  fields
defined
by Eq.\ref{simtrafo}
which are manifestly $\vartheta_1\leftrightarrow\vartheta_2$
symmetric.
To prove this
 we first observe that $\vartheta_1\leftrightarrow\vartheta_2$
takes $\varpi$ into $-\varpi$ (cf. Eq.\ref{zeromodes}), whereas  the
normalized operators $V_m^{(J)}$ behave as
\beq
V_m^{(J)} \quad \rightarrow \quad V_{-m}^{(J)}
\label{Uexchange}
\eeq
(similarly for the right-movers). The latter follows
by comparison of the conformal weights and zero mode shifts of
the  $V_m^{(J)}$ operators built from $\vartheta_1$ resp.
$\vartheta_2$,
as these two properties define normalized primary fields uniquely.
For positive half-integer $J$, the summation range
in Eq.\ref{Cbgas} is $m=-J,\dots J$, hence symmetric under $m \to
-m$,
and the exchange $\vartheta_1 \leftrightarrow \vartheta_2$
essentially
amounts only to a reorganization of terms. Then after commuting the
$T$ operators to the left or to the right on both sides,
Eq.\ref{sl2inv}
can be solved straightforwardly. A particular solution is
\beq
T(\varpi,\varpib)=\sqrt{
{\Gamma(1-\varpi h/\pi) \Gamma(1-\varpib h/\pi)\sqrt{\lfloor
\varpi\rfloor
\lfloor \varpib\rfloor} \over
\Gamma(1+\varpi)\Gamma(1+\varpib)}}\mu_0^{(\varpi+\varpib)/4}
\label{T}
\eeq
The last factor means effectively\footnote{Actually the last factor
in
Eq.\ref{T} removes only $\mu_0^m$ in Eq.\ref{muc}, but the remaining
normalization constant $\mu_0^J$ plays no role here. It will become
important,
 however, when we consider the equations of motion.} that we should
put
 $\mu_0=1$ in Eq.\ref{3.31}
and Eq.\ref{Cbgas} (cf Eq.\ref{muc}), and so we will take $\mu_0=1$
in
the
following.
The solution Eq.\ref{T} is unique up to the replacement
$T(\varpi,\varpib)
\to T(\varpi,\varpib)T_1 (\varpi,\varpib)$, with
\beq
{ T_1(\varpi,\varpib)T_1(-\varpi-2m,-\varpib -2m)\over
T_1(\varpi+2m,\varpib+2m) T_1(-\varpi,-\varpib)}=1
\label{Tfreedom}
\eeq
Unfortunately the case of continous $J$, Eq.\ref{sl2inv} is not so
easy to analyze, as the family of operators $V_m^{(J)}$ with
$J+m=0,1,2,\dots$
is no longer
invariant under the replacement $m \to -m$, and Eq.\ref{sl2inv}
becomes
highly nontrivial. We leave this problem for  future work and
will
restrict also in the next subsection to the case of positive
half-integer $J$.
\subsubsection{Hermiticity}
\label{hermiticity}
Another property of the Liouville exponentials that has not yet been
discussed is hermiticity. As was worked out by Gervais and Neveu a
long
time ago\cite{GN4}, the free fields possess the following behaviour
under hermitian conjugation (for brevity of notation we write only
the
left-movers explicitly):
\[
\vartheta_1^{{\dag}} =\vartheta_1 \quad , \quad \vartheta_2^{\dag}
=\vartheta_2
\qquad (\varpi=-\varpi^\ast)
\]
\beq
\vartheta_1^{\dag} =\vartheta_2 \quad , \quad \vartheta_2^{\dag}
=\vartheta_1
\qquad (\varpi=\varpi^\ast)
\label{thetaherm}
\eeq
The first case corresponds to the hyperbolic sector of the theory,
the second to the elliptic sector which we consider here.
Consequently, one has for the vertex operators
resp. screening charges:
\[
{V_{-J}^{(J)}}^{\dag} =V_{-J}^{(J)} \ , \ {V_J^{(J)}}^{\dag}
=V_J^{(J)}
\ ,
\ S_{(i)}^{\dag} =S_{(i)} \quad (\varpi=-\varpi^\ast)
\]
\beq
{V_{-J}^{(J)}}^{\dag} =V_{J}^{(J)} \ , \ {V_J^{(J)}}^{\dag}
=V_{-J}^{(J)} \ ,
\ S_{(i)}^{\dag} =S_{(i)} \quad (\varpi=\varpi^\ast)
\label{Uherm}
\eeq
where $S_{(i)}$ denotes the screening charge constructed from
$\vartheta_{i}$.
It is then immediate to show that in the elliptic sector,
\beq
(e^{-J\alpha_-\Phi }_{(1)})^{\dag}=e^{-J\alpha_-\Phi }_{(2)} \qquad
(\varpi=\varpi^\ast)
\label{expherm2}
\eeq
Our exponentials can formally be interpreted also in the hyperbolic
sector,
and fulfill there
\beq
(e^{-J\alpha_-\Phi }_{(i)})^{\dag}=e^{-J\alpha_-\Phi }_{(i)} \qquad
(\varpi=
-\varpi^\ast)
\label{expherm1}
\eeq
However, their locality properties are not entirely obvious in this
sector-cf.
below. Returning to the elliptic case, we note that
Eq.\ref{sl2inv} and Eq.\ref{expherm2} imply
\[
(Te_{(1)}^{-J\alpha_-\Phi}T^{-1})^{\dag} =
C (Te_{(1)}^{-J\alpha_-\Phi}T^{-1}) C^{-1} \ ,
\]
with\hfill
\beq
C(\varpi,\varpib)=T^{-1{\dag}}(\varpi,\varpib)T^{-1}(-\varpi,-\varpib)
\label{expherm3}
\eeq
Thus, hermiticity is realized only up to a similarity transformation.
In fact, in the elliptic sector there exists no similarity
transformation
 $T$ at all such that
$C$ becomes trivial, even if Eq.\ref{sl2inv} is not imposed. This
fact
was first observed\footnote{More precisely, it was pointed out
in the second of refs.\cite{LuS}
 that $e^{-\alpha_-\Phi/2}$ can be chosen hermitian resp.
antihermitian
in certain regions of $\varpi,\varpib$ space, but
$e^{-\alpha_-\Phi/2}$
cannot be consistently restricted to these regions.}
in ref.\cite{LuS} and later rediscovered in \cite{BP}.
 Nevertheless, the weaker
hermiticity property Eq.\ref{expherm3} serves almost the same purpose
as
``true'' hermiticity as far as correlators of the Liouville
exponentials
are concerned, as the similarity transformation $C$ cancels out up to
the contributions from the end points where $C$ resp. $C^{-1}$ hits
the
left resp. right vacuum. A more serious problem in the elliptic
sector,
also observed in ref.\cite{LuS}, is that the  Liouville
exponentials
possess no natural restriction to the subspace of positive norm
states,
given by the condition $|\varpi| <1+\pi/h , \ |\varpib|<1+\pi/h$.
For the coupling of $c<1$ matter to gravity, however, this problem
is irrelevant, as all negative norm states become
decoupled through the Virasoro conditions.

\subsection{The Liouville Field $\Phi$}
\subsubsection{Definition}
Having constructed Liouville exponentials with arbitrary continous
spins,
we can now define the Liouville field $\Phi$ itself by\cite{OW}
\beq
\alpha_- \Phi := -{d\over dJ}\left.
e^{\textstyle -J\alpha_-\Phi}\right |_{J=0}
\label{4.1}
\eeq
Though $\Phi$ is not really a primary field --- it is similar to the
stress-energy tensor in this respect ---
it is needed to verify the validity of canonical commutation
relations
and the quantum equations of motion.
Thus we expand Eq.\ref{Cbgas} near $J=0$. In this limit, the factor
$\prod_{k=1}^{J+m}
\lfloor 1+2J-k\rfloor\to {2Jh\over \sin h}\prod_{k=2}^{J+m}
\lfloor 1+2J-k\rfloor$ vanishes  except for $J+m=0$, and  the
exponential
tends to one as it should.
It then follows immediately that
$$
\Phi(\sigma, \tau)  =- (\vartheta_1(u) +\bar \vartheta_1(v) ) +
{2h\over \alpha_- \sin h }\times
$$
\beq
\sum_{n=1}^\infty  \tilde \mu_0^n {1 \over \lfloor n\rfloor}
\prod_{k=1}^n {1\over \lfloor \varpi+2n-k \rfloor \lfloor \varpi+k
\rfloor}
S(u)^n \Sb(v)^n.
\label{4.2}
\eeq
\subsubsection{Periodicity properties and singularity structure}
In the hyperbolic sector where $\varpi=\varpib$, the Liouville field
of Eq.\ref{4.2} is manifestly periodic. However, inspecting the
periodicity
behaviour of $\Phi$ in the elliptic sector with $\varpi \ne \varpib$,
we find
that
\beq
\alpha_-\Phi(\sigma+2\pi,\tau)=\alpha_-\Phi(\sigma,\tau)-2\pi i k
\label{period}
\eeq
where $k$ is the parameter appearing in Eq.\ref{condx}. The constant
is
entirely
produced by the free field contribution to $\Phi$, as the series in
screening
charges is periodic order by order (cf. Eq.\ref{3.2}).
Eq.\ref{period}
obviously calls for some explanation, as at least classically the
Liouville
field should be periodic by definition. We will carry out the
discussion classically,
but this will suffice to obtain a qualitative understanding of the
situation.
The essential point is that the definitions Eq.\ref{2.2} and
Eq.\ref{4.1},
though seemingly equivalent classically, actually differ slightly in
the
elliptic sector. If $2\sqrt\gamma \Phi(\sigma,\tau)$ is regular
everywhere
in $[0,2\pi]$, the two definitions clearly can differ only by a
constant. But
it is well known\cite{JKM} that in the elliptic sector (with $k\ne
0$)
there
are
$|k|$ nonintersecting singularity lines, thus $|k|$ singularities in
$\sigma \in
[0,2\pi]$. At a singular point, the constant connecting the two
definitions
may -  and does - change, creating in this way a nontrivial
periodicity
behaviour of our field $\Phi$ of Eqs.\ref{4.1}, \ref{4.2}. Indeed,
classically
the definition Eq.\ref{4.1} is equivalent to
\beq
2\sqrt\gamma\Phi=-2\ln {A'}^{-1/2}-2\ln B{B'}^{-1/2} -2\ln (1-A/B) +
\hbox{const}.
\label{classdefphi}
\eeq
with the series expansion representing the logarithm. It follows
easily
from the results of \cite{JKM} that at each singular point, $2\ln
(1-A/B)$
 - and hence $\Phi$ of Eq.\ref{classdefphi} - jumps by an imaginary
constant $-2\pi i \hbox{sgn} k$, whereas there is no such jump, of
course,
in Eq.\ref{2.2} which is by definition real. Hence,
\[
2\sqrt\gamma(\Phi_{Eq.\ref{4.1}}(\sigma+2\pi,\tau)
-\Phi_{Eq.\ref{4.1}}(\sigma,
\tau))
\]
\beq
-2\sqrt\gamma(\Phi_{Eq.\ref{2.2}}(\sigma+2\pi,\tau)-\Phi_{Eq.\ref{2.2}}
(\sigma,\tau)) =-2\pi ik
\label{phidiff}
\eeq
As the second difference is zero, we reproduce Eq.\ref{period}. Thus
the
Liouville field we are using differs from the "true" one only by a
constant
between any two singularity lines, but the constant changes at the
singularities. In particular, our $\Phi$ cannot be real everywhere in
$\sigma \in [0,2\pi]$. (This has nothing to do with the
nonhermiticity
of the exponentials noted in section \ref{hermiticity}, as the latter
is
independent of $\sigma$; indeed, for the exponentials with
half-integer
$J$,
the jumps play no role for the hermiticity behaviour). We remark that
the
periodicity behaviour of our Liouville field is actually quite
natural,
as the spectrum in the elliptic sector contains a winding number
($k$)
and
therefore looks like that of a compactified field.

When using the free field $\vartheta_2$ instead of $\vartheta_1$,
the periodicity behaviour of $\Phi$ of Eq.\ref{4.2} will be exactly
opposite.
One may think that the regions of convergence for the two (classical)
expansions, $|A/B|<1$ resp. $|B/A|<1$ are complementary and therefore
there is no contradiction. However, in contrast to the hyperbolic
sector we
have
$|A/B| \equiv 1$ in the elliptic sector, such that both series
expansions are
exactly {\it on} their circle of convergence, and in fact converge
there except
for the singularities at $A=B$. Thus we see that
$\vartheta_1\leftrightarrow
\vartheta_2$ invariance is broken
by the singularities in the elliptic sector. However, for the
exponentials
with half-integer $J$, we get the same periodicity behaviour for the
$\vartheta_1$
and the $\vartheta_2$ representation. Correspondingly, we were able
even
in the quantum case to construct these exponentials in a $\vartheta_1
\leftrightarrow
\vartheta_2$ invariant way.
For continous $J$, however, it is not obvious how this
invariance can be
restored.

\subsubsection{The case $k=0$}
A special consideration is required for the case where $\varpi$ is
real
(elliptic sector) but $k=0$ in Eq.\ref{condx}. It follows from the
work of \cite{JKM} that in this situation, there is no real Liouville
field
even classically. However, from the point of view of the locality
analysis,
the case $k=0$ is very natural and therefore we did not exclude it.
It is not hard to show that indeed more generally one needs to have
\beq
\varpi\varpib <0
\label{realitycond}
\eeq
classically in order to obtain real solutions of the Liouville
equation
with
positive cosmological constant. In the other case, one has $
2\sqrt\gamma
\hbox{Im}\Phi =\pm i\pi$, and so the real part of $\Phi$ solves
 the Liouville equation
with negative cosmological constant. As regards the singularity
structure,
$\sigma$ and $\tau$ essentially exchange their roles, and thus one
obtains
 timelike instead of spacelike singularity lines\cite{P}.
The number of singularities can in general be greater than $|k|$,
though
$|k|$ continues to characterize the periodicity behaviour of our
solution
Eq.\ref{4.2}. The explanation is that the additional singularities
always
come in pairs with opposite associated jumps $\pm 2\pi i$ of
$2\sqrt\gamma
\Phi$, so their effect is not seen in the overall periodicity
behaviour
of $\Phi$. In particular, for $k=0$ we now understand why the
Liouville
field
Eq.\ref{4.2} is periodic in spite of the possible presence of
singularities.
We stress also that in our analysis there is no restriction on the
sign
of
$\varpi\varpib$, thus we describe solutions of the Liouville equation
with
both signs of the cosmological constant.

\subsubsection{ The field equations}
For the considerations in this subsection and below, the similarity
transformations $T$ discussed above play no role and so we return to
the
representation Eq.\ref{4.2} resp. Eq.\ref{Cbgas} of the Liouville
exponentials. It is straightforward to derive from  Eq.\ref{4.2}
explicit expressions for the derivatives of $\Phi$, and one obtains:
$$
\partial_u \Phi= -\partial_u \vartheta_1 +{2ih\over \alpha_- \sin h}
\sum_{n=1}^\infty
{\tilde \mu_0^{n} \over \lfloor n\rfloor}c_n(\varpi)\times
$$
\beq
\prod_{k=1}^{n}
{1 \over
\lfloor \varpi+2n-k \rfloor \lfloor \varpi+k \rfloor}
U^{(-1)}_{n} \Sb^n,
\label{duphi}
\eeq
with\hfill
\beq
c_n(\varpi)=2\sin h q^{\varpi+1}\lfloor n\rfloor \lfloor
\varpi+n\rfloor
\label{4.7}
\eeq
and similarly for $\partial_v \Phi$.
In  Eq.\ref{duphi}  we observe the appearance of the coefficients of
the
expansion Eq.\ref{Cbgas} with $J=-1$.  Taking the crossed derivative
we
thus get
\beq
\partial_u \partial_v \Phi=-{\alpha_-\over 8}
e^{\textstyle \alpha_-\Phi} \ ,
\label{fequ}
\eeq
if we choose\hfill
\beq
\tilde \mu_0={1\over 32\pi \sin h}
\label{defmu0}
\eeq
Eq.\ref{fequ} is the quantum Liouville field equation associated
with an action given by Eq.\ref{2.1} with $2\sqrt{ \gamma}$ replaced
by
$\alpha_-$, and  with $\mu=1$.
In view of the recently shown equivalence of different
frameworks\cite{GS2},
we can directly compare this result with the one obtained in an
older  analysis
by Otto and Weigt\cite{OW}, and find agreement\footnote{
Note that the formula of Otto and Weigt quoted in ref.\cite{GS2}
needs
to be
multiplied by a factor $({\sin h \over h})^{2J}$ to be in accord with
the
equations
of motion. (This was already noticed in \cite{OW}).}. At this point
it should be noted that the main approaches within the operator
framework, those of refs.\cite{BCGT}, refs.\cite{OW}, and the present
one, have been shown in \cite{GS2} to be equivalent, precisely by
means
of a similarity transformation of the form Eq.\ref{simtrafo}.

\subsubsection{Equal-time commutation relations}
We now proceed to the canonical commutation relations.
It follows trivially from Eqs.\ref{loc},
\ref{4.1} that
\beq
[\Phi(\sigma,\tau),\Phi(\sigma',\tau)]=0.
\label{4.17}
\eeq
Next, by differentiating Eq.\ref{4.17} twice with respect to time
and using the equations of motion, we see that also
\beq
[\Pi(\sigma,\tau),\Pi(\sigma',\tau)]=0.
\label{4.18}
\eeq
where $\Pi(\sigma,\tau)$ is the canonical momentum,
\beq
\Pi(\sigma,\tau)={1\over 4\pi} \partial_\tau \Phi(\sigma,\tau)
\label{defPi}
\eeq
Furthermore we note that $[\Pi(\sigma,\tau), \Phi(\sigma',\tau)]$ can
be
nonvanishing only at $\sigma =\sigma'$, due to the fact the
$R$-matrices
for arbitrary spins   depend on $\tau,\sigma$ only via the step
functions
$\theta(u-u')$ resp. $\theta(v-v')$. On the other hand, the
contribution
of the free field parts of $\Pi$ and $\Phi$ gives precisely the
expected
result:
\beq
[\Pi(\sigma,\tau),\Phi(\sigma',\tau)]|_{\hbox{ free field}}
=-i\delta(\sigma-\sigma')
\label{4.19}
\eeq
By straightforward arguments \cite{GS3} one can show that the sum of
the other
contributions vanishes. Thus one finds for the full commutator,
\beq
[\Pi(\sigma,\tau),\Phi(\sigma',\tau)]=-i\delta(\sigma-\sigma')
\label{4.22}
\eeq
as expected, showing that the quantization scheme is indeed
canonical.

\section{The case of two screening charges.  }
\subsection{The braiding}
In the above analysis, we have used only one quantum deformation
parameter $h$, which is related to the semiclassical screening charge
$\alpha_-$ according to Eq.\ref{screeningcharge} and tends to zero in
the classical limit $C\to \infty$
according to Eq.\ref{defh}. However, it is of course well known e.g.
from
the minimal models that there exists another screening charge
$\alpha_+$
with analogous properties, except that it blows up in the classical
limit.
Correspondingly, there is a second deformation parameter $\hhat$ with
\beq
\alpha_+\equiv 2\sqrt{\hhat \over 2\pi}
\label{screeningchargehat}
\eeq
(cf. Eq.\ref{screeningcharge}),
and $\hhat$ is given by Eq.\ref{defh} with the other sign of the
square root.
The relation between $h$ and $\hhat$ is simply
\beq
\hhat  \equiv \pi ^2/h
\label{2.35}
\eeq
Thus we obtain another set of screening operators $\Shat$ resp.
primary fields  $\Uhat_{\mhat}^{(\Jhat)}$ which
have the same form Eqs.\ref{3.1},\ref{3.3} as $S$ resp. $U_m^{(J)}$,
but involve $\hhat$ instead of $h$. Hence their exchange algebra (and
also their fusion properties) are the same as that of the
``unhatted''
operators. More generally, one can combine the fields
 $\Uhat_{\mhat}^{(\Jhat)}$ and $U_m^{(J)}$ by fusion
to obtain the operators $U_{m\mhat}^{(J\Jhat)}$
 with the properties (cf. Eqs.\ref{2.16}, \ref{3.4})
$$
\Delta_{J,\Jhat}=\Delta_{J+\Jhat\pi/h}
$$
\beq
U_{m\mhat}^{(J\Jhat)}\varpi =(\varpi +2m+2\mhat \pi/h)
U_{m\mhat}^{(J\Jhat)}\ ,\ U_{m\mhat}^{(J\Jhat)}\varpihat
=({\widehat \varpi} +2\mhat +2m h/\pi)
U_{m\mhat}^{(J\Jhat)}
\label{2.36}
\eeq
where $\widehat \varpi =\varpi  h/\pi$.  For half-integer $J,\Jhat$
the braiding (and fusion) of the $U_{m\mhat}^{(J\Jhat)}$
follows immediately from that of the $U_m^{(J)}$ because the two sets
of operators $U_m^{(J)}$ and $\Uhat_\mhat^{(\Jhat)}$ commute up to a
phase:
\beq
U_m^{(J)}(\sigma) \Uhat_\mhat^{(\Jhat)}(\sigma')=e^{-2\pi iJ\Jhat
\epsilon(\sigma-\sigma')}\Uhat_\mhat^{(\Jhat)}(\sigma')
U_m^{(J)}(\sigma)
\label{2.37}
\eeq
The natural expectation is that this will remain true even for
noninteger $2J$. However, here we meet a surprise.
 The commutation of $V_{-J}^{(J)}
(\sigma)$ and $\Vhat_{-\Jhat}^{(J)}(\sigma')$ gives the factor in
Eq.\ref{2.37}, the screening charges $S(\sigma)$ and $\hat
S(\sigma')$
commute, but
\beq
S_{\sigma\sigma'} \Vhat_{-\Jhat}^{(\Jhat)}(\sigma')= e^{2\pi i\Jhat}
\Vhat_{-\Jhat}^{(\Jhat)}(\sigma')S_{\sigma\sigma'}
\label{2.38}
\eeq
whereas\hfill
\beq
S_\Delta \Vhat_{-\Jhat}^{(\Jhat)}(\sigma')= e^{-2\pi i\Jhat}\Vhat_{
-\Jhat}^{(\Jhat)}(\sigma')S_{\Delta }
\label{2.39}
\eeq
The phase factors agree only when $2J$ is integer, and thus the
commutation
of hatted and unhatted operators becomes nontrivial in general.
We should therefore restart the machinery of section 2 with the
operators $U_{m\mhat}^{(J\Jhat)}$, where
\beq
U_{m \mhat}^{(J\Jhat )}:= V_{-J}^{(J)}\Vhat_{-\Jhat}^{(\Jhat)}
S^{J+m}{\hat S}^{\Jhat +\mhat}
\label{2.40}
\eeq
with the product of the first two factors being defined by
renormalizing the
short-distance singularity  as usual (cf. also ref.\cite{GS1}).
One obtains the following result for the $R$-matrix of the
$U_{m\mhat}^{(J\Jhat)}$ :
\beq
 R_U(\underline{J},\underline{J}';\varpi)_{
\underline{m}_{\phantom {2}} \underline{m}'}^{\underline{m}_2
\underline{m}_1}=q^{-\Je{\Je}'}\qhat^{-\Jehat{\Jehat}'}
R_U(\Je,{\Je}';\varpi )_{\ms {\ms}'}^{\mstwo \msone}
\hat R_U(\Jehat,{\Jehat}';\varpihat )_{\mshat {\mshat}'}
^{\mstwohat \msonehat}
\label{2.43}
\eeq
Here, we have introduced "effective spins"
\beq
\Je:=J+\Jhat \pi/h  \quad , \quad \Jehat := \Jhat + Jh/\pi
\label{2.42}
\eeq
and shifted $m$'s
$$
\ms= m-\Jhat {\pi\over h},\quad \ms'= m'-\Jhat' {\pi\over h},\quad
\msone= m_1-\Jhat {\pi\over h},\quad \mstwo= m_2-\Jhat' {\pi\over h}
$$
\beq
\mshat= \mhat-J {h\over \pi},\quad \mshat'= \mhat-J' {h\over
\pi},\quad
\msonehat=\mhat_1-J {h\over \pi},\quad
\mstwohat= \mhat_2-J' {h\over \pi}.
\label{ms}
\eeq
Their appearance is easy to understand upon noting that
$V_{-J}^{(J)}\Vhat_{-\Jhat}^{(\Jhat)}\propto V_{-\Je}^{(\Je)}$.
So indeed everything can depend only on the effective spins and the
screening
numbers
$$
n=J+m=\Je +\ms,\quad n'=J'+m'=\Je'+\ms',
$$
\beq
\nhat=\Jhat+\mhat=\Jehat +\mshat,\quad \nhat'=\Jhat'+\mhat'=\Jehat +
\mshat'
\label{ndef}
\eeq
and similarly for $n_1, n_2, \nhat_1, \nhat_2$ (with $n+n'=n_1+n_2$,
$\nhat+\nhat'=\nhat_1+\nhat_2$).
The $R$-matrices $R_U$ and $\hat R_U$ in Eq.\ref{2.43}
are given by the same expression
Eq.\ref{3.20}, written in terms of the deformation parameter $h$
resp.
$\hhat$.

Thus in the continuous case
$2J$ and $2\Jhat$ loose meaning, since it is not possible
to recover them from $\Je$. Correspondingly, the fields
$U_{m\mhat}^{(J\Jhat)}$ should be thought of as related to
a new quantum group $U_{q\hat q}$ whose representations are labelled
by the effective spin $\Je$ and carry two magnetic quantum numbers
$\ms,\mshat$ instead of one. A more detailed elaboration of its
properties
is in preparation.
We can  write the complete $R$-matrix again in the
form  Eq.\ref{3.20}, if we introduce "vectorial"
quantities  $\kappa_{\Jgen1
,\Jgen2 }^{\Jgen{12} }$
and $\left\{ ^{\Jgen1 }_{\Jgen3 }\,^{\Jgen2 }_{\Jgen{123} } \right.
 \left |^{\Jgen{12} }_{\Jgen{23} }\right\}$ with $\Jgen{}
=(J,\Jhat)$
etc.
(cf. also ref.\cite{CGR1}).  Then
$$
R_U(\underline{J},\underline{J}';\varpi)_{
\underline{m}_{\phantom {2}} \underline{m}'}^{\underline{m}_2
\underline{m}_1}= e^{-i\pi(\Delta_{x} +\Delta_{x+m^e_{1}+m^e_{2}}
-\Delta_{x+m^e} -\Delta_{x+m^e_2})}
$$
$$
\times
{\kappa_{\Jgen{} ,\xgen{} +\mgen{12} }^{\xgen{} +\mgen2 }
\kappa_{\Jgen{} ',\xgen{} +\mgen2 }^{\xgen{} } \over
\kappa_{\Jgen{} ',\xgen{} +\mgen{12} }^{\xgen{} +\mgen{} }
\kappa_{\Jgen{} ,\xgen{} +\mgen{} }^{\xgen{} }}
\left\{ ^{\Jgen{} }_{\Jgen{} ' }\,^{\xgen +\mgen{12} }_{\xgen }
\right.
 \left |^{\xgen +\mgen2 }_{\xgen +\mgen{} }\right\}
$$
where\hfill
\[
\kappa_{\Jgen{} ,\xgen +\mgen{} }^{\xgen }=\kappa_{\Je,x+\ms}^x
\kappahat_{\Jehat,\xhat +\mshat}^\xhat
\]
and\hfill
\beq
\left\{ ^{\Jgen{} }_{\Jgen{} ' }\,^{\xgen +\mgen{12} }_{\xgen }
\right.
 \left |^{\xgen +\mgen2 }_{\xgen +\mgen{} }\right\}=
\left \{^ {\Je }_{{\Je}' }\,^{x+\mssum}_{x} \right.
\left | ^{x+\mstwo }_{x +\ms }\right\}
\Gaghat\,   ^{\Jehat }_{{\Jehat}' }\, \,^{\xhat +\mssumhat }_{\xhat }
\bigr. \bverthat\, ^{\xhat +\mstwohat }_{\xhat +\mshat }  \Gadhat,
\label{2.44}
\eeq
where we have let $\mssum=\msone+\mstwo$,
$\mssumhat=\msonehat+\mstwohat$ and $m^e =m+\mhat \pi/h$.
We note that Eq.\ref{2.44} agrees precisely with the ``vectorial''
$6j$ symbol
found in ref.\cite{GR} where extensions of the usual $6j$ are
discussed
on the basis of the polynomial equations.
This completes  the derivation of the
braiding of the chiral vertex operators with both
screening charges. We add as a final remark that the generalized
$6j$ symbols fulfill orthogonality relations completely analogous
to those known for the usual ones. They take the form
\beq
\sum_{\Jgen{23} }
\left\{ ^{\Jgen{1} }_{\Jgen{3} }\,^{\Jgen{2} }_{\Jgen{123} }
\right.
 \left |^{\Jgen{12} }_{\Jgen{23} }\right\}
\left\{ ^{\Jgen{1} }_{\Jgen{3} }\,^{\Jgen{2} }_{\Jgen{123} }
\right.
 \left |^{\Kgen{12} }_{\Jgen{23} }\right\}=\delta_{\Jgen{12},
\Kgen{12}
}.
\label{orthogene}
\eeq
where the summation is over all $\Je_{23}$ such that
$\Je_2+\Je_3-\Je_{23}
\in {\bf Z}_+ +\pi/h {\bf Z}_+$.

\subsection{The generalized Liouville field}
The generalization of the Liouville exponential  is given by
\beq
e^{\textstyle -J^e\alpha_-{\underline \Phi}(\sigma, \tau )}=
\sum _{n, \nhat =0}^{\infty} \mu_0^{n} \muhat_0^{\nhat}
\Vt^{(\Je)}_{n \nhat}(u)
{\overline \Vt^{(\Je)}_{n \nhat}}(v) \>
\label{genexp}
\eeq
where $\muhat_0$ is given by equations  similar to Eqs.\ref{mu0tilde}
and \ref{defmu0} with $h\to \hhat$. Of course, the preference of
$\alpha_-$
over $\alpha_+$ in Eq.\ref{genexp} is purely notational as
$e^{-J^e\alpha_-{\underline \Phi}}
\equiv e^{-\Jhat^e \alpha_+{\underline \Phi}}$. Using
Eq.\ref{orthogene}, it is straightforward  to verify that the
generalized
exponential  is local
and closed by fusion, provided $\varpi=\varpib$. The previous
discussions of the hermiticity and Weyl reflection properties can
be easily generalized. We now turn to the generalized
Liouville field, which we
define again by Eq.\ref{4.1}. Thus we obtain
$$
{\underline \Phi}(\sigma, \tau)  =- (\vartheta_1(u) +\bar
\vartheta_1(v) )
$$
$$
+{2h\over \alpha_- \sin h }
\sum_{n=1}^\infty  \tilde \mu_0^n {1 \over \lfloor n\rfloor}
\prod_{k=1}^n {1\over \lfloor \varpi+2 n-k \rfloor \lfloor \varpi+k
\rfloor}
S(u)^n \Sb(v)^n.
$$
\beq
+{2\hhat\over \alpha_+ \sin \hhat }
\sum_{\nhat=1}^\infty  \tilde \muhat_0^n {1
\over \lfloorhat \nhat\rfloorhat}
\prod_{\hat k =1}^\nhat {1\over \lfloorhat \varpihat+2\nhat-\hat k
\rfloorhat
 \lfloorhat \varpihat +\hat k \rfloorhat}
\Shat(u)^{\nhat} \Shatb(v)^{\nhat}.
\label{genLiou}
\eeq
It differs from the previous one by the last line. Since
$\varpi=\varpib$,
$\Phi$ is periodic.
As a result, the
quantum field equation becomes
\beq
\partial_u \partial_v {\underline\Phi}=-{\alpha_-\over 8}
e^{\textstyle \alpha_-\Phi} -{\alpha_+\over 8}
e^{\textstyle \alpha_+\hat\Phi},
\label{genfequ}
\eeq
involving both  cosmological terms . Since ${\underline
\Phi}=\Phi+\hat\Phi
+\vartheta_1+\bar \vartheta_1$, the validity of Eq.\ref{genfequ} is a
trivial
consequence of the equations of motion with a single screening
charge.
The reason why ${\underline \Phi}$ must be shifted w.r.t. $\Phi+\hat
\Phi$ is
that
 $\Phi+\hat \Phi$ alone is not local. Indeed, using
Eqs.\ref{kappashift} and
its
left-moving analog, and the fact that screening charges of different
type
commute,
we see that the only nonzero contributions to
$[\Phi(\sigma,\tau)+\hat\Phi(\sigma,\tau),
\Phi(\sigma',\tau)+\hat\Phi(\sigma',\tau)]=[\Phi(\sigma,\tau),
\hat\Phi(\sigma',\tau)]+
[\hat\Phi(\sigma,\tau),\Phi(\sigma',\tau)]$ are of the form
$[\vartheta_1,\Shat]$ resp.
$[\vartheta_1,S]$. These commutators are precisely cancelled by the
free field
shift.
Finally, one may easily extend the previous discussion of the
canonical
commutation relations. The result is that Eqs.\ref{4.17}, \ref{4.18}
and
\ref{4.22}
remain true  for the generalized Liouville field without any
modification. If we regard the passage from $h$ to $\hhat$, or
$\alpha_-$ to
$\alpha_+$, as some kind of duality transformation, then
Eq.\ref{genfequ}
describes the selfdual version of the theory. It would be interesting
to
find a Lagrangian description of this theory -which is of course
not simply given by the usual Liouville Lagrangian  with both
cosmological
terms, as should be clear from the remarks above.

\section{Group-theoretic approach to the
classical Liouville exponentials}
We now switch subjects completely and turn to the question of the
quantum
$\tau$ function.
As a preparation of the coming section, it is pedagogical
to  temporarily return  to the classical case. We will consider only
the case of half-integer positive spins explicitly. Then we
have\footnote{In this section, we redefine the Liouville field, for
simplicity, so that the coupling constant need not be written any
more.}
the general solution of ref.\cite{LS}:
$$
e^{\textstyle -j\Phi(z,\, \zb)}=
{<j,\, j| \Mb^{-1}(\zb) M(z) |j,\, j>\over (s(z) \sb(\zb))^j}
$$
\beq
{d M\over dz}=s(z) M j_-,\quad
{d \Mb\over d\zb}=\sb(\zb) \Mb j_+,
\label{6.1}
\eeq
where $s$ and $\sb$ are arbitrary functions of  a single variable.
The symbols  $j_\pm$ represent  $sl(2)$ generators satisfying
$[ j_+,\, j_-]=2 j_3$, and $|j,\, j>$ are highest-weight
states $j_+|j,j> =0$.  In Eq.\ref{6.1} and below, we use Euclidean
coordinates on the sphere,
$z=e^{\tau +i\sigma }$, $\zb =e^{\tau -i\sigma }$.
To establish the connection with the approach of the preceding
sections,
note first that $s,\sb$ are related to the arbitrary functions
$A,B$
appearing in the general classical solution Eq.\ref{2.2} by
\beq
s(z)=A'(z), \qquad \sb (\zb)=-B'(\zb)/B^2
\label{6.2}
\eeq
and thus $s$ and $\sb $ can be identified with the classical
equivalents of the screening charge densities $V_1^{(-1)}$, $\bar
V_1^{(-1)}$
(up to normalization). It is then immediate to verify that
Eq.\ref{6.1}
reduces to Eq.\ref{2.5} after evaluation of the
matrix element. Actually, the $SL(2,{\bf C})$ symmetry  Eq.\ref{2.3}
allows us just as well to identify $s^{-1/2},\sb ^{-1/2}$ with any
linear
combination of the quasiperiodic (Bloch wave) fields
$V_{-1/2}^{(1/2)}$ and $V_{+1/2}^{(1/2)}$ resp. $\bar
V_{-1/2}^{(1/2)}$
and $\bar V_{+1/2}^{(1/2)}$. For the classical considerations below,
the assignment chosen is irrelevant; however, on the quantum level
the
situation will be quite different.
Looking at Eq.\ref{6.1}, one may wonder
why  the Liouville exponential should be given by highest-weight
matrix elements. The basic reason is that
Eq.\ref{6.1} must be such that
\beq
e^{\textstyle -j_1\Phi(z,\, \zb)} e^{\textstyle -j_2\Phi(z,\, \zb)}
=e^{\textstyle -(j_1+j_2)\Phi(z,\, \zb)},
\label{4.3}
\eeq
by the very definition of the classical exponential function.
In order to verify this, we introduce the rescaled
exponentials\footnote
{For the conceptual considerations here, the precise normalization of
the
$E^{(j)}$ is not important. Therefore we don't specify the lower
integration
limits
in Eq.\ref{4.4a}.}
$$
E^{(j)}(s,j_-)=s^{-j}(z)M(z)= s^{-j}(z)e^{\int^z s(z')dz'
\ {\displaystyle
j_-}}
$$
resp.\hfill
\beq
\bar E^{(j)}(\sb ,j_+)=\sb ^{-j}(\zb)\bar M^{-1}(\zb)=
\sb ^{-j}(\zb)e^{-\int^\zb \sb (\zb')d\zb' \ {\displaystyle j_+}}
\label{4.4a}
\eeq
Thus the left-hand side of Eq.\ref{4.3} can be written as
$$
e^{\textstyle -j_1\Phi(z,\, \zb)} e^{\textstyle -j_2\Phi(z,\, \zb)}=
(<j_1,\, j_1| <j_2,\, j_2|)\bar E^{(j_1)}(\sb ,j_+) \otimes
\bar E^{(j_2)}(\sb ,j_+)\times
$$
$$
E^{(j_1)}(s,j_-) \otimes
E^{(j_2)}(s,j_-)(|j_1,\, j_1>|j_2,\, j_2>)
$$
$$
= (<j_1,\, j_1| <j_2,\, j_2|)\bar E^{(j_1+j_2)}(\sb ,j_+\otimes 1
+1 \otimes j_+)\times
$$
\beq
E^{(j_1+j_2)}(s,j_- \otimes 1 +1 \otimes j_-)
(|j_1,\, j_1>|j_2,\, j_2>)
\label{4.4}
\eeq
The highest weight-states considered are the only ones such that
the tensor product  gives a single  irreducible representation.
Its  spin is $j_1+j_2$, and  $|j_1,\, j_1>|j_2,\, j_2>$ is
the highest-weight vector. Since the matrix element of
$\bar M^{-1} M$ is  determined solely  by  the group structure, it
only
depends upon the spin of the representation, and not upon the way it
is
realized;
 hence Eq.\ref{4.3} follows.
In particular, we have
\beq
e^{\textstyle -j\Phi(z,\, \zb)}=
\left (e^{\textstyle -(1/2)\Phi(z,\, \zb)}\right)^{2j};
\label{4.5}
\eeq
and thus Eq.\ref{6.1} may be re-written using binomial coefficients
(more on this below).

It is well-known (see e.g. ref.\cite{GM})
that the Liouville solution and its Toda generalization
are actually tau-functions  in the sense of the Kyoto group\cite{JM}.
A characteristic feature of tau-functions is to involve
 highest-weight states. We shall not dwell into the precise
connection, since it is not directly evident from Eq.\ref{6.1}.
We shall rather recall  the existence of bilinear equations of the
Hirota type, which was the original motivation to introduce tau
functions.
The method of derivation we will use
 is not the same as the standard ones of
ref.\cite{JM} or ref.\cite{KW}. Its interest is that
 it will remain applicable in the
quantum case.

 One may obtain  a closed equation for the $j=1/2$
Liouville exponential as follows. Making use of
Eq.\ref{6.1}, let us compute  the antisymmetric
bilinear expression
$$
e^{ -\Phi/2} \partial_z \partial_\zb
e^{-\Phi/2} -
\partial_z e^{ -\Phi/2} \partial_\zb
e^{-\Phi/2}=
$$
$$
-<{1\over 2},\, {1\over 2} | \Mb^{-1} M |{1\over 2},\, {1\over 2}>
<{1\over 2},\, -{1\over 2} | \Mb^{-1} M |{1\over 2},\, -{1\over 2}>
$$
\beq
+<{1\over 2},\, -{1\over 2} | \Mb^{-1} M |{1\over 2},\, {1\over 2}>
<{1\over 2},\,  {1\over 2} | \Mb^{-1} M |{1\over 2},\, -{1\over 2}>
\label{4.5a}
\eeq
The point of this particular combination of
derivatives is that the functions $s$ and $\sb$ disappear, and
the result
is given by the matrix element of $\Mcb^{-1} \Mc$
in the $j=0$ representation, where
$\Mc=\Mcb=1$. Thus we get
\beq
e^{ -\Phi/2} \partial_z \partial_\zb
e^{-\Phi/2} -
\partial_z e^{ -\Phi/2} \partial_\zb
e^{-\Phi/2}=-1.
\label{4.6}
\eeq
Of course it is trivial to rederive this equation directly from the
Liouville equation; however
this form  -- the simplest example of
Hirota bilinear equations -- which only makes use of the
Liouville exponentials and not of the field $\Phi$ itself,  will
be much
easier to generalise to the quantum case. For later use we note
that this Hirota equation is equivalent to the following relation in
the
Taylor expansion  for $z'\to z$, $\zb'\to \zb$.
\beq
e^{ -\Phi/2( z',  \zb')}
e^{ -\Phi/2(z ,\zb)} -
e^{ -\Phi/2( z',  \zb)}
e^{ -\Phi/2(z ,\zb')}\sim -(z'-z)(\zb'-\zb)
\label{6.7}
\eeq
Clearly we may obtain other bilinear equations for $\exp(-j \Phi)$
with $j\not=1/2$ by again projecting out the $j=0$ component of the
product.

\section{Liouville exponentials and q-deformations}

\subsection{The quantum group structure}
The deep connection of the quantum Liouville theory
with $U_q(sl(2))$ was there from the
beginning\cite{GN4}, but in disguise.
It was elucidated more recently in refs.\cite{B,G1,CG2,CGR1,CGR2}.
Though the fusing
and braiding matrices  of the $V_m^{(J)}$ fields are given in terms
of
quantum group ($6j$) symbols, they are not quantum group covariant;
there
exists another basis of chiral vertex operators $\xi_M^{(J)}$,
related to the $V_m^{(J)}$
by a linear transformation, which behave as spin $J$ representations
of $U_q(sl(2))$. The transformation takes the form
\cite{G1}\cite{G3}\cite{CGR2}
\beq
\xi_M^{(J)}(\sigma)= |J,\varpi)_M^m E_m^{(J)}(\varpi) V_m^{(J)}
\label{jomtrafo}
\eeq
where $|J,\varpi)_M^m$ and $E_m^{(J)}(\varpi)$ are suitable
transformation
coefficients, whose explicit values are not important for our
purposes (they
can be found in refs.\cite{G1}\cite{G3}\cite{CGR2}, for the case of
half-integer
positive spins which we consider here).
Their fusing and braiding matrices are given by
q Clebsch-Gordan and universal R-matrix elements respectively.
In particular, the operator-product algebra of the $\xi_M^{(J)}$
corresponds to
making
q tensor products of representations.
We anticipate therefore that it is this basis which should be used
when
trying to extend the considerations of the previous section to the
quantum level, so that the arbitrariness mentioned below Eq.\ref{6.2}
in the
precise assignment of $s(z),\sb (\zb)$ is lifted.
It was shown in ref.\cite{G5}
that  the Liouville
exponentials take the form ($2J$ positive integer)
\begin{equation}
e^{\textstyle -J\alpha_-\Phi(z, \zb  )}=
\sum _{M=-J}^J\> (-1)^{J+M}  \>e^{ih(J+M)}\>
\xi_M^{(J)}(z)\,
{\overline \xi_{-M}^{(J)}}(\zb)
\label{5.1}
\end{equation}
The point of this section is to show that with  this last expression
for the Liouville exponential,  {\bf its quantum
properties are  directly connected with their
 classical analogues  by  standard q deformations}.

\subsection{Connection with q-binomials}

We shall use the notations of refs.\cite{G1},\cite{CG2}--\cite{CGR2}.
One
introduces
\beq
\lfloor x \rfloor = {\sin (h x)\over \sin(h)}.
\label{5.2}
\eeq
The q-deformed binomial coefficients
 noted  ${P \choose Q}$ are defined by
\beq
{P \choose Q} := {\lfloor P \rfloor \! !
 \over \lfloor Q \rfloor \! !\lfloor P-Q \rfloor \! !},
\qquad \lfloor n \rfloor \! ! :=
\prod_{r=1}^n \lfloor r \rfloor,
\label{5.3}
\eeq
They are binomial coefficients for
 the expansion of $(x+y)^{2J}$, with
$x$ and $y$ non-commuting variables such that
\beq
xy=yx e^{-2ih}.
\label{5.4}
\eeq
 Indeed, it
is easy to verify that they satisfy
\beq
{m+1 \choose n}=e^{ihn}{m \choose n}+e^{-ih(m-n+1)}{m \choose n-1}.
\label{5.5}
\end{equation}
As a result one  sees that if one lets
\beq
(x+y)^N= \sum_{r=0}^N {N\choose r} e^{ ihr(N-r)} x^r y^{N-r}
\label{5.6}
\end{equation}
one has, as required,
\beq
(x+y)^{N+1}=(x+y) (x+y)^{N}.
\label{5.7}
\eeq

We now use the leading-order fusion properties of the $\xi$
fields\cite{G1}:
to leading order in the
short distance singularity at  $z \to z'$,
the product of $\xi$ fields behaves as\footnote{We are working
here on the sphere, rather than on the cylinder.}
\beq
\xi_M^{(J)}(z)\,\xi_{M'}^{(J')}(z')
\sim
(z-z')^{-2JJ'h/ \pi}
\>\lambda (J,M;\,J',M')
\,\xi_{M+M'}^{(J+J')}(z' ),
\label{5.8}
\eeq
\beq
 \lambda (J,M;\,J',M')=
\sqrt{ {  {2J \choose J+M}}\>
{ {2J' \choose J'+M'}}
\over {  {2J+2J' \choose J+J'+M+M'}} }
\>e^{ih(M'J-MJ')}.
\label{5.9}
\eeq
Thus, if we redefine
\beq
\eta_M^{(J)}\equiv \xi_M^{(J)}/ \sqrt{{2J \choose J+M}}
\label{5.10}
\eeq
we have
$$
\eta_M^{(J)}\,\eta_{M'}^{(J')}
\sim
\eta _{M+M'}^{(J+J')} \>e^{ih(M'J-MJ')}
$$
and thus \hfill
\beq
 \eta _M^{(J)}\sim (\eta _{1/2}^{(1/2)})^{J+M}(\eta
_{-1/2}^{(1/2)})^{J-M}
e^{{ih\over 2}(J^2-M^2)}
\label{5.11}
\eeq
In the above formulae and hereafter the symbol $\sim$ means
leading term of the short-distance expansion, divided by the
singular short distance factor appearing in Eq.\ref{5.8}.
In terms of the
$\eta$ fields, the Liouville exponential takes the form

\beq
e^{\textstyle -J\alpha_-\Phi(z, \zb )}
=\qquad \sum _{J+M=0}^{2J} {2J \choose J+M} (-1)^{J+M}
e^{ih(J+M)}\eta_M^{(J)}
(z)\etab _{-M}^{(J)}(\zb )
\label{5.12}
\eeq
or, using Eq.\ref{5.11},\hfill
$$
e^{\textstyle -J\alpha_-\Phi(z, \zb )} \sim \sum _{J+M=0}^{2J}
{2J \choose J+M} e^{ih(J+M)(J-M)}\times
$$
\beq
\left (-e^{ih}\eta_{1/2}^{(1/2)}\,
{\overline \eta_{-1/2}^{(1/2)}}\right )^{J+M}
\left (\eta_{-1/2}^{(1/2)}\,
{\overline \eta_{1/2}^{(1/2)}}\right )^{J-M}
\nnn
\label{5.13}
\eeq
which is completely analogous to the q-binomial expansion
Eq.\ref{5.6},
if we identify $x=-e^{ih}\eta_{1/2}^{(1/2)}(z)\,{\overline
\eta_{-1/2}^{(1/2)}}
(\zb )\, ,
\  y=\eta_{-1/2}^{(1/2)}(z)\, {\overline \eta_{1/2}^{(1/2)}}(\zb )$.
To avoid any possible confusion, we remark that Eq.\ref{5.4} does not
imply
that the fields $\eta_{1/2}^{(1/2)}(z)\,{\overline
\eta_{-1/2}^{(1/2)}}(\zb )$
and $\eta_{-1/2}^{(1/2)}(z')\, {\overline \eta_{1/2}^{(1/2)}}(\zb '
)$ commute
up to a factor; this is true only in the limit $z'\to z,\ \zb' \to
\zb$.
 We remark
that, in view of the above construction of arbitrary Liouville
exponentials,
Eq.\ref{5.12} can easily be generalized to continous spins; however,
this is
not necessary for the points we want to make here.

\subsection{Liouville exponentials as quantum tau-functions}
As is well known,
the binomial coefficients  are closely related to representations of
$U_q(sl(2))$.  Consider group-theoretic states\footnote{We assume
for simplicity that $h/\pi$ is not rational.}
$\vert J, M >$, $-J\leq M\leq J$;
 together with operators $J_{\pm}$, $J_3$ such that:
\beq
J_\pm \vert J,M> =\sqrt{\lfloor J \mp M\rfloor
\lfloor J \pm M+1 \rfloor } \vert J, M\pm 1 >
\quad J_3 \vert J,M> =M\, \vert J,M>.
\label{5.15}
\eeq
These operators satisfy the $U_q(sl(2))$  commutation relations
\beq
\Bigl[J_+,J_-\Bigr]=\lfloor 2J_3 \rfloor, \quad
\Bigl[J_3,J_\pm \Bigr]=\pm J_\pm.
\label{5.16}
\eeq
It is elementary to derive the formulae
$$
<J,\, N\vert  (J_+)^P \vert J,M> =\sqrt{ \lfloor J+N\rfloor \! !
\lfloor J-M \rfloor \! ! \over \lfloor J-N\rfloor \! !
\lfloor J+M \rfloor \! !} \delta_{N,\, M+P}
$$
\beq
<J,\, N\vert  (J_-)^P \vert J,M> =\sqrt{ \lfloor J-N\rfloor \! !
\lfloor J+M \rfloor \! ! \over \lfloor J+N\rfloor \! !
\lfloor J-M \rfloor \! !} \delta_{N,\, M-P}
\label{5.17}
\eeq
Recall further that the co-products  of representations are  defined
by
\beq
\Lambda (J)_\pm=J_\pm\otimes e^{ihJ_3}+
e^{-ihJ_3}\otimes J_\pm,\quad
\Lambda (J)_3=J_3\otimes 1+1\otimes J_3
\label{5.20}
\eeq
We will show now that the group-theoretical
classical formulae  of section 4 possess direct
quantum equivalents, obtained
by replacing $sl(2)$ by
$U_q(sl(2))$. We  start  from the general
classical solution and consider Eq.\ref{6.1} as the classical
tau function. The quantum tau function should then be given by a
representation of type Eq.\ref{6.1} of the operator $ e^{\textstyle
-J\alpha_-\Phi}$; that is, the q tau function should actually
be an operator instead of a function.
Let us introduce the (rescaled) q-exponentials
$$
E_q^{(J)}(\eta (z),J_-)=\sum_{J+M=0}^\infty \eta_M^{(J)}
{(J_-)^{J+M}\over \lfloor J+M\rfloor !}
$$
\beq
\bar E_q^{(J)}(\etab (\zb),J_+)=\sum_{J+M=0}^\infty e^{ih(J+M)}
(-1)^{J+M} \etab_{-M}^{(J)}
{(J_+)^{J+M}\over \lfloor J+M\rfloor !}
\label{5.22}
\eeq
which we take to be the quantum equivalents of the classical
(rescaled)
exponentials $E^{(j)}(s,j_-)$, $\bar E^{(j)}(\sb ,j_+)$.
Indeed, in the
limit $h\to 0$, we see that $E_q^{(J)}(\eta ,J_-)$,
$\bar E_q^{(J)}(\etab ,J_+)$ reduce to $E^{(j)}(s,j_-)$,
$\bar E^{(j)}(\sb,j_+)$ if
we identify classically
$$
\eta_M^{(J)} =s^{-J}(z)(\int^z s(z')dz')^{J+M}=
\eta_{-J}^{(J)}(\eta_1^{(0)})^{J+M}=
\eta_{-J}^{(J)} \eta_{J+M}^{(0)}
$$
\beq
\etab_{-M}^{(J)} =\sb ^{-J}(\zb )(\int^\zb \sb (\zb')d\zb')^{J+M}
=\etab_{J}^{(J)}
(\etab_{-1}^{(0)})^{J+M}=
\etab_{J}^{(J)} \etab_{-J-M}^{(0)},
\eeq
which corresponds to the assignment
\beq
s(z)=\eta_1^{(-1)}(z) \qquad \sb (\zb )=\etab_{-1}^{(-1)}(\zb )
\eeq
Furthermore we will show now, following closely the calculation used
to derive
Eq.\ref{4.3} group-theoretically, that indeed
the $E_q^{(J)}$ obey a composition law appropriate for q-deformed
exponentials.
The argument was inspired by ref.\cite{GKL} (with an important
difference-
see below).
In ref.\cite{CGR2}, the fusion algebra of the $\xi$ fields was
determined
using the general scheme of Moore and Seiberg. One has
$$
\xi ^{(J_1)}_{M_1}(z_1)\,\xi^{(J_2)}_{M_2}(z_2) =
\sum _{J_{12}= \vert J_1 - J_2 \vert} ^{J_1+J_2}
g _{J_1J_2}^{J_{12}} (J_1,M_1;J_2,M_2\vert J_{12})\times
$$
\beq
\sum _{\{\nu\}} \xi ^{(J_{12},\{\nu\})} _{M_1+M_2}(z_2)
<\! <\!\varpi _{J_{12}},{\{\nu\}} \vert
V ^{(J_1)}_{J_2-J_{12}}
(z_1-z_2) \vert \varpi_{J_2}\! > \! >,
\label{5.29}
\eeq
where $\{\nu\}$ is a multi-index that labels
the  descendants, and $|\varpi _{J},{\{\nu\}} \! > \! >$ denotes
the corresponding state in the Virasoro Verma-module.
Similar equations hold for the $\xib$ fields.
The explicit expression of the  coupling constant $g$ is not needed
in the present argument. The symbol $(J_1,M_1;J_2,M_2\vert J_{12})$
denotes the q-Clebsch-Gordan coefficients.  It follows
from their  very definition (see, e.g. ref.\cite{G3})  that
$$
\sum_{M_1+M_2=M_{12}} (J_1,M_1;J_2,M_2\vert J_{12})
|J_1,\, M_1>\otimes |J_2,\, M_2>=
$$
\beq
\phantom{\sum_{M_1+M_2=M_{12}} (J_1,M_1;J_2,M_2\vert J_{12})
|J_1,\, M_1>}=
 {(\Lambda(J)_- )^{J_{12}-M_{12}}\over
\lfloor J_{12}-M_{12}\rfloor \! ! \sqrt{{2J_{12}\choose
J_{12}-M_{12}}}}
 |J_{12},\, J_{12}>,
\label{5.30}
\eeq
and one finds
$$
E_q^{(J_1)}\! \left (\eta(z_1),\,
J_-\otimes 1
 \right )
E_q^{(J_2)}\! \left (\eta(z_2),\,
1\otimes  J_-\right )
(\vert J_1,\, J_1 >\vert J_2,\, J_2 >)=
$$
$$
\sum _{J_{12}= \vert J_1 - J_2 \vert} ^{J_1+J_2}
g _{J_1J_2}^{J_{12}} \times
$$
\beq
\sum _{\{\nu\}} E_q^{(J_{12} ,\{\nu\})}\! \left (\eta(z_2),\,
\Lambda(J)_- \right )|J_{12},\, J_{12}>
<\! <\!\varpi _{J_{12}},{\{\nu\}} \vert
V ^{(J_1)}_{J_2-J_{12}}
(z_1-z_2) \vert \varpi_{J_2}\! > \! >
\label{5.31}
\eeq
In particular, to leading order one has
$$
E_q^{(J_1)}\left (\eta(z_1),\,
J_-\otimes 1
 \right )
E_q^{( J_2)}\left (\eta(z_2),\,
1\otimes  J_-\right )
(\vert J_1,\, J_1 >\vert J_2,\, J_2 >)\sim
$$
\beq
E_q^{(J_{1}+J_2 )}\left (\eta(z_2),\,
\Lambda(J)_- \right )|J_{1}+J_2,\, J_{1}+J_2>.
\label{5.32}
\eeq
which is the natural multiplication law for
q-exponentials involving quantum group generators.
The coproduct is non-symmetric between the two
representations. On the left hand side this
comes from the non-commutativity   of the $\eta $ fields as quantum
field operators. Thus the present definition  of q exponentials
is conceptually
different from the usual one where the group ``parameters''
are c numbers. Let us recall the latter for completeness. If on
defines
\beq
e_q(X)\equiv \sum_{r=0}^\infty
 {X^r\over \lfloor r\rfloor\! !} e^{-ihr(r+1)/2},
\label{x.17}
\eeq
one has
\beq
e_q(x J_\pm \otimes e^{ihJ_3})
e_q(x e^{-ihJ_3} \otimes J_\pm )= e_q(x \Lambda (J)_\pm).
\label{x.20}
\eeq
Since they transform the q sum of infinitesimal generators into
products, the q exponentials are the natural  way to exponentiate
q Lie algebras. The last equation should be compared with
Eq.\ref{5.32}.
Now the non-symmetry of the co-product is taken care of by
exponentiating non-commuting group elements ($J_\pm \otimes
e^{ihJ_3}$,
and $e^{-ihJ_3} \otimes J_\pm$ on the left-hand side), with x a
number.
This is in contrast with Eq.\ref{5.32}, where $J_\pm \otimes 1$, and
$1 \otimes J_\pm$ are used instead.

Finally, we can rewrite Eq.\ref{5.12} under the
form\footnote{Related formulae are already given in ref.\cite{LS2},
p. 27.}
\begin{equation}
e^{\textstyle -J\alpha_-\Phi(z, \zb )}=
<J,\, J \vert \bar E_q^{(J)}(\etab(\zb ),\,   J_+)
E_q^{(J)}(\eta(z),\, J_-) \vert J,\, J>=\tau_q (\eta ,\bar\eta ).
\label{5.25}
\end{equation}
Indeed we will see that it
 possesses the obvious q-analogues of the properties Eq.\ref{4.4}
and Eq.\ref{4.5}. Hence,
Eq.\ref{5.25} should be viewed as the quantum version of the
Leznov-Saveliev
formula Eq.\ref{6.1}.
Since $\eta$ and $\etab$ commute, we deduce that
$$
e^{\textstyle -J_1\alpha_-\Phi(z_1, \zb_1 )}
e^{\textstyle -J_2\alpha_-\Phi(z_2, \zb_2 )}=
$$
$$
(<J_1,\, J_1 \vert <J_2,\, J_2 \vert)
 \bar E_q^{(J_1)}\! \left (\etab(\zb_1),\,
J_+\otimes 1\right)
\bar E_q^{(J_2)}\! \left (\etab(\zb_2),\,
1\otimes  J_+\right  )
$$
\beq
E_q^{(J_1)}\! \left (\eta(z_1),\,
J_-\otimes 1
 \right )
E_q^{(J_2)}\! \left (\eta(z_2),\,
1\otimes  J_-\right )
(\vert J_1,\, J_1 >\vert J_2,\, J_2 >).
\label{5.28}
\eeq

Making use of the fusion algebra Eq.\ref{5.29}, together with its
counterpart for the $\bar \eta$ fields gives back the fusion algebra
of the Liouville exponentials derived in ref.\cite{G5}. To leading
order one has
\beq
e^{\textstyle -J_1\alpha_-\Phi}
e^{\textstyle -J_2\alpha_-\Phi}\sim
e^{\textstyle -(J_1+J_2)\alpha_-\Phi}.
\label{zzz}
\eeq
Clearly these properties are natural generalizations of the
classical features recalled in section 6.

\subsection{The quantum Hirota equation}
At present, we cannot yet write the direct quantum equivalent of
Eq.\ref{4.6}.
Instead, we will characterize the way in which the information of the
quantum
Hirota equation is encoded into the theory. For this purpose, we
return
to Eq.\ref{6.7}, which is certainly equivalent to the classical
Hirota equation
Eq.\ref{4.6}. On the other hand, Eq.\ref{6.7} should possess meaning
on the quantum level, if the role of  the Taylor expansion is taken
over  by the operator product.
The OPE of the
$J=1/2$ Liouville
exponential  has the form\cite{G5}
$$
e^{\textstyle -{1\over 2}\alpha_-\Phi(z_1, \zb_1 )}
e^{\textstyle -{1\over 2}\alpha_-\Phi(z_2, \zb_2 )} \sim
\left[ (z_1-z_2)(\zb_1-\zb_2)\right ] ^{-h/2\pi}
 e^{\textstyle -\alpha_-\Phi(z_2, \zb_2 )}
$$
\beq
+\left[(z_1-z_2)(\zb_1 -\zb_2) \right ] ^{1+3h/2\pi}
c_0+\> \hbox{descendants},
\label{5.33}
\eeq
where $c_0$ is a constant that can be changed by a global shift
of the Liouville field.
The second Liouville exponential is equal to a constant, since its
spin
$J$ is equal to zero. This property is the quantum equivalent of
Eq.\ref{6.7}. Thus, the quantum Hirota equation should be equivalent
to the fact that
for $J=0$, $\exp( -J\alpha_-\Phi)=\hbox{ cst}$.
One sees that choosing a particular combination of derivatives
in a bilinear classical expression
of $J=1/2$
Liouville exponentials  is replaced by  picking up the spin zero
term in the operator-product expansion of $\exp( -(1/2)\alpha_-\Phi)$
with itself.
Note that due to the
quantum effects, the difference of the powers of
$(z_1-z_2)(\zb_1 -\zb_2)$
between the first and second term is not equal to one --- it is
equal to $1+2h$ ---   so that a simple antisymmetrization is not
enough,
as in the classical case,  to remove the first term.
Clearly, Eq.\ref{5.33} gives non-trivial equations relating the
matrix elements of the quantum Liouville exponentials.

\section{Conclusions/Outlook}
The operator approach to Liouville Theory, which originated more than
ten years ago, has come a long way. Starting from the analysis of the
simplest Liouville field - the inverse square root of the metric -
which
corresponds to the $J=1/2$ representation, it has now progressed
to the construction of the most general Liouville operators in the
standard (weak coupling) regime, corresponding to arbitrary
highest/lowest weight representations of the quantum group.
The underlying chiral algebra, either in its Bloch wave/Coulomb gas
or its quantum group covariant guise, has revealed beautiful
structures
which may find applications also in very different contexts.
 Though the completion
of the quantization program of Gervais and Neveu thus finally comes
into sight
for the weak coupling sector,  there remains an important complex of
questions yet to be addressed. While the Coulomb gas picture
presented
here leads immediately to integral representations of arbitrary
n-point functions in the half-integer positive spin case\cite{LuS},
the correlators of operators with continous spins require more care.
This is due to the fact that outside the half-integer positive $J$
region,
the sums
representing the Liouville exponentials become infinite, and their
evaluation
within correlation functions is quite nontrivial even in the simplest
case
of the three-point function\cite{Cargese}.
In ref.\cite{G5}, where the computation
of three-point functions relevant for minimal matter coupled to
gravity
was
discussed (here $J$ is half-integer {\it negative}) , these
difficulties were
avoided
in an interesting way by using instead of the "canonical" expression
for the
Liouville exponential as given by Eq.\ref{genexp}
another operator with the same conformal
weight, which is however represented by a {\it finite} sum.
This approach appears
to be closer in spirit to the analytic continuation
procedure employed in the
path integral framework, and its connection
with the first-principle approach
along the lines
of this review certainly deserves a better understanding.

The techniques used in this work are applicable not only
to the standard weak coupling ($C>25$) regime considered here,
but also to the strong coupling theory as developed in refs.\cite{G3}
\cite{GR}. In this case, however, it turns out that one needs
inverse powers of the screening charges Eq.\ref{3.1}. These can be
immediately formulated in the Gervais-Neveu framework, as the
replacement $A\to -1/A$ which inverts the screening charge just
corresponds to exchanging the free field $\vartheta_1$ with
$\vartheta_2$. Using the polynomial equations, and the fact that
$S(\sigma) S^{-1}(\sigma)=1$ (without renormalization) it is actually
possible to relate the braiding of
negative powers of the screenings to that of positive powers, and one
can rigorously control the exchange algebra resp. locality properties
of the strong coupling operators. (This will be explained elsewhere
 in detail).

In the last section, we have been discussing a possible approach
to the problem of defining a $\tau$ function for the quantum
Liouville
theory, thus embedding our treatment into the general framework of
integrable systems. We have seen that the classical Leznov-Saveliev
formula has an immediate and very natural quantum generalization.
The emerging quantum $\tau$ function, due to its operator
nature, is a noncommutative object and this is compatible with the
interesting results of \cite{GKLMM}\footnote{there, and in
ref.\cite{MMV},
also the concept of a commuting $\tau$ function is discussed},
 which however on the quantum level are at present still restricted
to a subset of solutions of the full dynamics.
On the other hand, our analysis is also still somewhat incomplete,
as we have not yet derived the quantum Hirota equation in a
satisfactory
form. Given the fact that our classical Hirota equation is directly
equivalent
to the Liouville equation, we expect that  the quantum Hirota
equation must be related in a similar way to the (known) quantum
Liouville
equation . From that point of view, one would actually not expect
that
the differential equation becomes replaced by a q-difference equation
on the
quantum level, as in \cite{GKLMM}. Certainly, these questions deserve
further investigation.

\vskip 5mm
\noindent
{\bf Acknowledgements}

We are grateful to E. Cremmer, D. Lebedev,  J.-F. Roussel and G.
Weigt for useful discussions.

\appendix
\section{ Some  previous results.}
\label{previous results}
This section is meant as a brief account of some previous reults, for
the
reader who has already some familiarity with the present approach.
For lack of space, we refer the reader to refs.\cite{G1},\cite{CGR1}
for details on the conventions used.
In refs. \cite{B,G1}the quantum group structure was
shown
to be
of the type $U_q(sl(2))\odot U_\qhat (sl(2))$ for half-integer
positive spins, where $h$ is given by
Eq.\ref{defh}, and
\beq
\hhat={\pi \over 12}\Bigl(C-13
+\sqrt {(C-25)(C-1)}\Bigr),
\label{2.17}
\eeq
Each quantum group parameter is associated with a screening charge by
the
relations  $h=\pi (\alpha_-)^2/2$, $\hhat=\pi (\alpha_+)^2/2$.
The basic family of $(r,s)$ chiral operators in 2D gravity may be
labelled
by two quantum group spins $J$ and $\Jhat$, with
$r=2\Jhat+1$, $s=2J+1$, so that
the spectrum of
Virasoro weights is given by
\beq
 \Delta_{J\Jhat}
={C-1\over 24}-
{ 1 \over 24} \left((J+\Jhat+1) \sqrt{C-1}
-(J-\Jhat) \sqrt{C-25} \right)^2,
\label{2.18}
\eeq
in agreement with Kac's formula.
One outcome of ref.\cite{CGR1} was the fusion and braiding of
the general chiral operators $V_{\mge}^{(\Jge )}$, also denoted
$V_{m \mhat}^{(J \Jhat)}$, where underlined  symbols denote double
indices
$\Jge \equiv (J,  \, \Jhat)$, $\mge \equiv (m,\,  \mhat)$,  which
were
all taken to be half integers:
$$
{\cal P}_{\Jgen{} } V_{\Jgen23 -\Jgen{} }
^{(\Jgen1 )}
V_{\Jgen3 -\Jgen23 }^{(\Jgen2 )} =
\sum_{\Jgen12 }
{g_{\Jgen1 \Jgen2 }^{\Jgen{12} }
\ g_{\Jgen{12} \Jgen3 }^{\Jgen{} }
\over
g _{\Jgen2 \Jgen3 }^{\Jgen{23} }
\ g_{{\Jgen1 }\Jgen{23} }^{\Jgen{} }
}
\left\{
^{\Jgen1 }_{\Jgen3 }\,^{\Jgen2 }_{\Jgen{} }
\right. \left |^{\Jgen{12} }_{\Jgen{23} }\right\} \times
$$
\beq
{\cal P}_{\Jgen{} } \sum_{\{\nu\}}
V_{\Jgen3 -\Jgen{} } ^{(\Jgen12 ,\{\nu\} )}
<\!\varpi _\Jgen12 ,\{\nu\}  \vert
V ^{(\Jgen1 )}_{\Jgen2 -\Jgen12 } \vert \varpi_{\Jgen2 }\! >.
\label{2.19}
\eeq
$$
{\cal P}_{\Jgen{} } V_{\Jgen23 -\Jgen{} }
^{(\Jgen1 )}
V_{\Jgen3 -\Jgen23 }^{(\Jgen2 )} =
\sum_{\Jgen13 }
e^{\pm i\pi (\Delta_\Jgen{} +\Delta_\Jgen3
-\Delta_{\Jgen23 }-\Delta_{\Jgen13 })}\times
$$
\beq
{g_{\Jgen1 \Jgen3 }^{\Jgen{13} }
\ g_{\Jgen{13} \Jgen3 }^{\Jgen{} }
\over
g _{\Jgen2 \Jgen3 }^{\Jgen{23} }
\ g_{{\Jgen1 }\Jgen{23} }^{\Jgen{} }
}
\left\{
^{\Jgen1 }_{\Jgen2 }\,^{\Jgen3 }_{\Jgen{} }
\right. \left |^{\Jgen{12} }_{\Jgen{23} }\right\}
{\cal P}_{\Jgen{} }
V_{\Jgen13 -\Jgen{} } ^{(\Jgen2 )}
V_{\Jgen3 -\Jgen13 }^{(\Jgen1 )},
\label{2.20}
\eeq
In these  formulae,  world-sheet variables are omitted,
and $\varpi$ is the rescaled zero-mode momentum of $\vartheta_1$ as
in
Eq.\ref{defomega}. It characterizes the Verma modules
${\cal H}(\varpi)$, spanned by states noted $|\varpi, \, \{\nu\}>$,
where
$\{\nu\}$ is a multi-index. In the generic case, where the Verma
module is trivial, ${\cal H}(\varpi)$ is a Fock space generated by
the non-zero modes of the
free field $\vartheta_1$ (or equivalently of $\vartheta_2$),
with the  ground state
 $|\varpi>$.
  The symbol $\varpi_{\Jge}$ stands for
$\varpi_0+2J+2\Jhat \pi/h$ where $\varpi_0 =1+\pi/h$, and
 ${\cal P}_{\Jgen{} }$ is the projector on
${\cal H}(\varpi_{\Jge})$. The above formulae contain the recoupling
coefficients for the
quantum group structure $U_q(sl(2))\odot U_\qhat(sl(2))$, which
are defined by
\beq
\left\{
^{\Jgen1 }_{\Jgen3 }\,^{\Jgen2 }_{\Jgen{} }
\right. \left |^{\Jgen{12} }_{\Jgen{23} }\right\}
=
(-1)^{ \fusV 1,2,,23,3,12 }
\left\{
^{J_1}_{J_3}\,^{J_2}_{J}
\right. \left |^{J_{12}}_{J_{23}}\right\}
\gaghat
\,^{\Jhat_1}_{\Jhat_3}\,^{\Jhat_2}_{\Jhat}
\bigr. \bverthat \, ^{\Jhat_{12}}_{\Jhat_{23}}\gadhat
\label{2.22}
\eeq
where $\left\{
^{J_1}_{J_3}\,^{J_2}_{J}
\right. \left |^{J_{12}}_{J_{23}}\right\} $ is the $6j$ coefficient
associated with  $U_q(sl(2))$, while $\gaghat
\,^{\Jhat_1}_{\Jhat_3}\,^{\Jhat_2}_{\Jhat}
\bigr. \bverthat \, ^{\Jhat_{12}}_{\Jhat_{23}}\gadhat$ stands for the
$6j$  associated with $U_{\qhat}(sl(2))$. $\fusV 1,2,,23,3,12 $ is an
integer given by
\beq
\fusV 1,2,123,23,3,12
=2\Jhat_2(J_{12}+J_{23}-J_2-J_{123})
+2J_2(\Jhat_{12}+\Jhat_{23}-\Jhat_2-\Jhat_{123})
\label{sign}.
\eeq

In addition to these group
theoretic features  there appear the coupling constants
        $g_{{\Jgen{1} }\Jgen{2} }^{\Jgen{12} }$ whose expression
was given in ref.\cite{CGR1}. In order to connect with the
 general setting recalled in section \ref{self-contained},
let us indicate that $V^{(J, 0)}_{-J, 0}$ is proportional  to
$f_{-J}^{J}|_{\hbox {\scriptsize qu}}$,
and that $V^{(J, 0)}_{J, 0}$ corresponds to the normal
ordered exponential of $\vartheta_2$, that is to
$f_{J}^{J}|_{\hbox {\scriptsize qu}}$. One may
verify --- this is left as an exercise to the dedicated reader ---
that the equations just written are such that the braiding of these
fields
is simply
\beqa
V^{(J_1, 0)}_{-J_1, 0}(\sigma_1)\>   V^{(J_2, 0)}_{-J_2,
0}(\sigma_2)&=&
e^{-2ih J_1 J_2\epsilon(\sigma_1-\sigma_2)}
 V^{(J_2, 0)}_{-J_2, 0}(\sigma_2)\>
 V^{(J_1, 0)}_{-J_1, 0}(\sigma_1), \label{2.23} \\
V^{(J_1, 0)}_{J_1, 0}(\sigma_1)\>  V^{(J_2, 0)}_{J_2, 0}(\sigma_2)&=&
e^{-2ih J_1 J_2\epsilon(\sigma_1-\sigma_2)}
V^{(J_2, 0)}_{J_2, 0}(\sigma_2)\>
 V^{(J_1, 0)}_{J_1, 0}(\sigma_1).
\label{2.24}
\eeqa
where $\epsilon(\sigma_1-\sigma_2)$ is the sign of
$\sigma_1-\sigma_2$
(for definiteness, we consider the interval  $0\leq \sigma_i\leq
2\pi$).
This confirms that they are normal ordered exponentials  of
free fields, in agreement with the starting point of the
GN quantization.  Note that the braiding of $V^{(J, 0)}_{-J, 0}$ with
$V^{(J, 0)}_{J, 0}$ involves  the full complexity of the
6-j coefficients, so that the commutation relations of $\vartheta_1$
with $\vartheta_2$ are definitely not of   free-field type.


\begin{thebibliography}{99}
\bibitem{B} O. Babelon,
\pl   B215, 1988, 523 .
\bibitem{G1} J.-L. Gervais,  \cmp 130, 1990, 257 .
\bibitem{G2} J.-L. Gervais, \pl B243, 1990, 85 .
\bibitem{CG2} E. Cremmer, J.-L. Gervais,
 \cmp 144, 1992, 279 .
\bibitem{G4} J.-L. Gervais, \ijmp 6, 1991, 2805 .
\bibitem{G3} J.-L. Gervais, \cmp 138, 1991, 301 .
\bibitem{G5} J.-L. Gervais, ``Quantum group derivation
of 2D gravity-matter coupling'' Invited talk at
the Stony Brook meeting {\sl String and Symmetry 1991},
\np B391, 1993, 287 .
\bibitem{CGR1} E. Cremmer, J.-L. Gervais,
J.-F. Roussel, \np B413, 1994, 244.
\bibitem{CGR2} E. Cremmer, J.-L. Gervais,
J.-F. Roussel, \cmp 161, 1994, 597.
\bibitem{B2} O. Babelon,
\cmp 139, 1991, 619.
\bibitem{GN4} J.-L. Gervais, A. Neveu, \np B238, 1984, 125;
ibid., p. 396.
\bibitem {GS1} J.-L. Gervais, J. Schnittger,
\pl B315, 1993, 258 \hfill
\bibitem{BPZ} A.A. Belavin, A.M. Polyakov, A.B. Zamolodchikov,
\np B241, 1984, 332.
\bibitem{DF}  Vl. Dotsenko, V. Fateev,
\np B251, 691, (1985) .
\bibitem{MS} G. Moore, N. Seiberg, \cmp 123, 1989, 77.
\bibitem{GM}   J.-L. Gervais, Y. Matsuo,
\pl B274, 1992, 309; \cmp 152, 1993, 317.
\hfill
\bibitem{GR} J.-L. Gervais, J.-F. Roussel, \np B426, 1994, 140.
\bibitem{ANPS} A. Anderson,
B.E.W. Nilsson, C.N. Pope and K.S.Stelle,
\np B430, 1994, 107.\hfill
\bibitem{GN3} J.-L. Gervais, A. Neveu,  \np B224, 1983, 329.
\bibitem{GN2} J.-L. Gervais, A. Neveu,  \np B209, 1982, 125.
\bibitem {LuS} D. L\"ust, J. Schnittger,
\ijmp A6, 1991, 3625; J. Schnittger,  Ph.D. thesis,
 Munich 1990.\hfill
\bibitem{JKM} L. Johannson, A. Kihlberg, R. Marnelius,
\prd 29, 1984, 2798;
L. Johannson, R. Marnelius,
\np B254, 1985, 201.\hfill
\bibitem{JM} see, e. g.  M. Jimbo, T. Miwa,  {\sl Publications of the
R.I.M.S.} vol {\bf 19}, No 3, 1983. \hfill
\bibitem{KW} V. Ka\v c, M. Wakimoto, {\sl Proc. Symp. Pure
Math.} {\bf 49} (1989) 191. \hfill
\bibitem{GS2} J.-L. Gervais, J. Schnittger,
\np B413,  1994, 277.
\bibitem{GS3} J.-L. Gervais, J. Schnittger,
\np B431, 1994, 273.
\bibitem{OW} H.J. Otto, G. Weigt,
\pl B159, 1985, 341; {\sl Z. Phys. } {\bf C31},
(1986) 219; G. Weigt, ``Critical exponents of conformal fields
coupled to two-dimensional quantum gravity in the conformal gauge'',
talk given at 1989 Karpacz Winter School of Theor. Physics, preprint
PHE-90-15; G. Weigt, ``Canonical quantization of the Liouville
theory,
quantum group structures, and correlation functions'', talk given at
1992
Johns Hopkins Workshop on Current Problems in Particle Theory,
Goteborg,
Sweden, hep-th 9208075, print-92-0383 (DESY-IFH).\hfill
\bibitem{BCGT} E.Braaten, T.Curtright, C.Thorn,
\pl B118, 1982, 115;
\prl  48, 1982, 1309;
\annp 147, 1983, 365;
 E.Braaten, T.Curtright, G.Ghandour, C.Thorn,
\prl 51, 1983, 19;
\annp 153, 1984, 147.\hfill
\bibitem{P} G. Jorjadze, A. Pogrebkov, M. Polivanov,
{\sl Teor. Mat. Fiz.}  {\bf 40} (1979) 221;
A. Pogrebkov, {\sl Teor. Mat. Fiz.} {\bf 45} (1980) 161;
G. Jorjadze, A. Pogrebkov, M. Polivanov, S. Talalov,
{\sl J. Phys. A: Math. Gen.}  {\bf 19} (1986) 121.
\bibitem{BP} J. Balog and L. Palla,
\pl B274, 1992, 232.
\bibitem{Cargese} J. Schnittger, proceedings of the
1993 NATO adv. research workshop
on new developments in string theory,
conformal models and
field theory, Cargese May 93,
Plenum Press, to appear.\hfill
\bibitem{DF}  Vl. Dotsenko, V. Fateev,
\np B251, 1985, 691 .\hfill
\bibitem{Fe} G. Felder, \np B317, 1989, 215; erratum \np B324, 1989,
548.\hfill
\bibitem{LS}
A.~N.~Leznov, M.~V.~Saveliev, \pl B79, 1978, 294;
{\sl Lett. Math. Phys.} {\bf 3} (1979) 207;
\cmp 74, 1980, 111;
Lett. Math. Phys. {\bf 6} (1982) 505;
\cmp 89, 1983, 59;
{\it Group-Theoretical Methods for Integration
of Nonlinear Dynamical Systems.} Progress
in Physics v.~15, Birkhauser-Verlag,
1992.\hfill
\bibitem{LS2}
A.~N.~Leznov, M.~V.~Saveliev, {\sl Acta Applicandae Mathematicae}
{\bf 16}
(1989) 1. \hfill
\bibitem{KS}
K. Kajiwara, J. Satsuma, {\sl J. Phys. Soc. Jpn.}, {\bf 60} (1991)
3986;
K. Kajiwara, Y. Ohta, J. Satsuma, "q-Discrete Toda Molecule
Equation",
 solv-int/9304001.\hfill
\bibitem{SMI}
F. Smirnov, "What are we quantizing in integrable field theory",
hep-th/9307097.\hfill
\bibitem{GKL} A. Gerasimov, S. Khoroshkin and D. Lebedev,
"q-deformations of $\tau$ functions: a toy model",
talk presented at the P. Cartier seminar at Ecole Normale Superieure,
Paris,
1993.
\bibitem{GKLMM}
A. Gerasimov, S. Khoroshkin, D. Lebedev, A. Mironov, A. Morozov,
ITEP-M-2-94, hep-th 9405011;
A. Mironov, FIAN-TD-12-94, hep-th 9409190. \hfill
\bibitem{KM}
S. Kharchev and A. Mironov, "$\tau$ function of two-dimensional Toda
lattice
and quantum deformations of integrable hierarchies", FIAN-TD-02-93.
\bibitem{MMV}
A. Mironov, A. Morozov, L. Vinet,
FIAN-TD-22-93, hep-th 9312213.\hfill


\end{thebibliography}
\end{document}